\documentclass[twocolumn]{aastex62}%

\newcommand{\lya}{Ly$\alpha$}

\newcommand{\halpha}{\mbox{H$\alpha$}}
\newcommand{\hbeta}{\mbox{H$\beta$}}

\newcommand{\hI}{\mbox{H{\sc i}}}

\newcommand{\oI}{\mbox{O{\sc i}}}

\newcommand{\oIII}{\mbox{O{\sc iii}}}

\newcommand{\cII}{\mbox{C{\sc ii}}}

\newcommand{\siII}{\mbox{Si{\sc ii}}}

\newcommand{\lstar}{\mbox{$L^\star$}}
\newcommand{\phistar}{\mbox{$\phi^\star$}}

\usepackage[utf8]{inputenc}
\graphicspath{{./}{figures/}}

\shorttitle{LARS X: Multivariate Lya prediction}
\shortauthors{Runnholm et al.}

\begin{document}

\title{LYMAN ALPHA REFERENCE SAMPLE: X. PREDICTING LYMAN ALPHA OUTPUT FROM STARFORMING GALAXIES USING MULTIVARIATE REGRESSION \footnote{Based on observations made with the NASA/ESA Hubble Space Telescope, obtained at the Space Telescope Science Institute, which is operated by the Association of Universities for Research in Astronomy, Inc., under NASA contract NAS 5-26555. These observations are associated with program \#12310, 12583, 13438, 13654.}}

\correspondingauthor{Axel Runnholm}
\email{axel.runnholm@astro.su.se}

\author[0000-0002-1025-7569]{Axel Runnholm}
\affil{Department of Astronomy, Oscar Klein Centre, Stockholm University, AlbaNova universitetscentrum, SE-106 91 Stockholm, Sweden}

\author{Matthew Hayes}
\affil{Department of Astronomy, Oscar Klein Centre, Stockholm University, AlbaNova universitetscentrum, SE-106 91 Stockholm, Sweden}

\author{Jens Melinder}
\affil{Department of Astronomy, Oscar Klein Centre, Stockholm University, AlbaNova universitetscentrum, SE-106 91 Stockholm, Sweden}

\author{Emil Rivera-Thorsen}
\affil{Department of Astronomy, Oscar Klein Centre, Stockholm University, AlbaNova universitetscentrum, SE-106 91 Stockholm, Sweden}

\author{Göran Östlin}
\affil{Department of Astronomy, Oscar Klein Centre, Stockholm University, AlbaNova universitetscentrum, SE-106 91 Stockholm, Sweden}

\author{John Cannon}
\affiliation{Department of Physics \& Astronomy, Macalester College, 1600 Grand Avenue, Saint Paul, MN 55105, USA}
\author{Daniel Kunth}
\affiliation{Institut d’Astrophysique, Paris, 98 bis Boulevard Arago, F-75014 Paris, France}

\begin{abstract}
Understanding the production and escape of Lyman $\alpha$ (\lya{}) radiation from star-forming galaxies is a long standing problem in astrophysics.
The ability to predict the \lya{} luminosity of galaxies would open up new ways of exploring the Epoch of Reionization (EoR), and  to estimate \lya\ emission from galaxies in cosmological simulations where radiative transfer calculations cannot be done.
We apply multivariate regression methods to the Lyman Alpha Reference Sample dataset to obtain a relation between the galaxy properties and the emitted \lya{}.
The derived relation predicts the \lya{} luminosity of our galaxy sample to good accuracy, regardless of whether we consider only direct observables (root-mean-square (RMS) dispersion around the relation of $\sim 0.19$ dex) or derived physical quantities (RMS $\sim 0.27$ dex). We confirm the predictive ability on a separate sample of compact star-forming galaxies and find that the prediction works well, but that aperture effects on measured \lya\ luminosity may be important, depending on the redshift of the galaxy.
We apply statistical feature selection techniques to determine an order of importance of the variables in our dataset, enabling future observations to be optimized for predictive ability.
When using physical variables, we are able to determine that the most important predictive parameters are, in order, star formation rate, dust extinction, compactness and the gas covering fraction.
We discuss the application of our results in terms of studying the EoR and intensity mapping experiments.

\end{abstract}

\keywords{Starburst galaxies (1570), Lyman-alpha galaxies (978), Multivariate analysis (1913)}


\section{Introduction} \label{sec:intro}

A Lyman alpha (Ly$\alpha$) photon is created in $\sim 68\%$ of all ionized hydrogen recombinations, assuming case-B recombination \citep{dijkstra2014}. This makes it not only the brightest spectral line of hydrogen but also, in principle, a strong tracer of the ionized interstellar medium (ISM) of galaxies especially at high redshifts (z) where the line is redshifted into the optical and easily observable with large groundbased telescopes. 

This simple picture is complicated by the fact that \lya{} is a resonant line, meaning that it can be absorbed and re-emitted by neutral hydrogen. The cross-section for such a scattering event is very large, and a \lya{} photon at line center experiences an optical depth of one already at a column density $N_{HI}$ of $10^{14}$ cm$^{-2}$.
A \lya{} photon will thus undergo a significant number of scatterings in virtually any medium \citep{osterbrock1962,adams1972}, and should therefore be trapped inside galaxies, slowly diffusing out to large radii before escaping. However, thermal and bulk motions of the gas, as well as random wing scattering events can impart Doppler shifts to the photons' frequency\citep{dijkstra2014}. Frequency-shifted photons will no longer be in resonance with the bulk of the neutral hydrogen and can escape almost freely \citep{dijkstra2014,kunth1998}. At the same time there is a significant probability of dust absorption during the scattering process that depends on the exact distribution of \hI{} and dust \citep{neufeld1990,laursen2013a,duval2014} which further complicates the process.
From all of these considerations it is clear that the escape of \lya{}  from the inner regions of a galaxy through the neutral component of the interstellar medium is a complex, non-local, radiative transfer (RT) problem and that interpreting \lya{} emission from a given galaxy is a non-trivial task, see \citet{hayes2015} for a review.

While RT makes \lya{} intrinsically very complicated, it also means that the emergent flux and equivalent width could potentially encode information about the conditions in, and the distribution of, the scattering media the radiation has passed through. This means that \lya{} has the potential to be used as a probe of neutral gas at crucial phases in the history of the Universe, such as the Epoch of Reionization (EoR).
\citet{haiman1999} suggest that an observed decrease in number densities of \lya{} emitting galaxies at high redshift could be used to probe the progression of reionization.
Early realizations of this idea by \citet{malhotra2004,malhotra2006,kashikawa2006}, using \lya{} luminosity functions, were able place some constraints on reionization. \citet{dijkstra2007} however, noted the need to account for the evolving halo mass function in such studies.

Later studies used the fraction of galaxies that show \lya{} in emission ($X_{\mathrm{Ly}\alpha}$), the distribution of their equivalent widths,  and measures of the global escape fraction of \lya{} \citep{stark2010,fontana2010,hayes2011,stark2011,pentericci2011,ono2012,treu2012,caruana2012,caruana2014,schenker2014,jung2018} to further demonstrate that \lya{} emission decreases at $z>6.5$, possibly as a consequence of increasing neutral hydrogen fraction in the IGM. \citet{mason2018a,mason2019} used Bayesian  methods to further refine this approach and constrain the neutral fraction at $z=7$ to $\simeq0.6$ and at $z=8$ to $> 0.76$.

These studies all use the evolution of relatively simple population statistics to determine when the IGM neutral fraction starts increasing. Such statistics circumvent the complicated issues of \lya{} escape from galaxies but can only provide a very generalized picture of galaxy evolution and the EoR. These statistics also implicitly assume that the intrinsic properties of the galaxy population do not significantly change between redshift 6, where IGM absorption is negligible, and redshift $\gtrsim 7$. This assumption is uncertain since halo mass functions, gas fractions, star formation histories, metallicities and more are all expected to change with redshift.  If the intrinsic \lya{} emission from galaxies could be predicted from properties whose observability is unaffected by neutral \hI, it would be possible to take the changing properties of the galaxies into account and differences between observed and predicted \lya\ emission for individual galaxies could be directly attributed to the neutral IGM. Using such an approach, we could therefore potentially study the IGM neutral fraction during the later stages of the EoR in much greater detail. Developing such a predictive relation is the focus of this paper. 

Two studies have previously derived relations of this kind.\citet{yang2017} construct a bivariate relation using the dust extinction and the velocity of the red \lya{} peak and find an dispersion of approximately 0.3 dex. However, this prediction includes a parameter directly measured from the \lya\ line which may limit its applicability in cases when \lya\ is not detected, or the signal-to-noise is insufficient to properly determine the red peak velocity. \citet{trainor2019} present a bivariate relation for \lya\ prediction. They do not report an RMS of the resulting relation which makes comparison to \citet{yang2017} or our work difficult (see Section~\ref{sec: Trainor}) but it does show some significant predictive ability. In this paper we want to fully explore the potential of multivariate analysis to determine how well \lya{} can be predicted when we have access to a large number of independent galaxy properties and if it is possible to use the data themselves to select which variables are the most important and need to be included in the fit.

We use data from  the Lyman Alpha Reference Sample (LARS) \citep{hayes2013,ostlin2014} to construct multivariate linear regressions that can accurately predict the \lya{} properties of the sample. We divide the data set into observational properties (such as far UV and H$\alpha$ luminosity) and derived physical quantities (such as stellar mass and SFR). The reason for doing so is that both observational and theoretical studies need to predict \lya\ but they have access to different qantities.

The LARS survey was specifically designed to facilitate this kind of work by covering all the parameters that are believed to significantly impact the escape of \lya{}.
We use variable selection methods to establish an order of importance of the variables we use for the prediction. This kind of technique might yield new insights into the RT problem and could also provide a method for prioritization of observations of high redshift galaxies with future facilities such as \emph{James Webb Space Telescope} (JWST), and 30m class groundbased telescopes for \lya\ prediction. We perform extensive tests of the predictive ability of our final relation both by using statistical cross-validation techniques and by using our derived relation to predict the \lya\ of a completely separate set of Green Pea and Lyman Break Analog galaxies and we find that our results are robust and generalize well to predictions outside our sample.

This paper is structured as follows. 
The full dataset that we use is presented in Section 2 and the methods used to obtain the prediction and evaluate the predictive ability and variable importance are explained in Section 3.
The results are presented in Section 4. In Sections 5 and 6 we discuss the implications of our results for future high redshift surveys and the outlooks for extending the current work.


\section{Sample and Data}\label{sec: Data}
\subsection{Observations}
The main set of observations used in this paper is part of the LARS and eLARS projects and in its current state presents an unprecedented wealth of multi--wavelength data related to the escape and physics of Ly$\alpha$.  
LARS is a sample of galaxies specifically selected to provide sufficiently detailed observations to shed light on these complicated processes. 
The 14 original galaxies were selected to be close by,  $0.028 \leq z \leq 0.18$, and highly star forming (by having H$\alpha$ equivalent width (EW) $\geq$ 100\AA\ and GALEX FUV luminosity $9.5 \leq \log(\nu L_\nu / L_\odot) \leq 10.7$). The sample was later extended with a further 28 galaxies where the H$\alpha$ EW limit was dropped to 30\AA\ (the eLARS sample) and the resulting H$\alpha$ EW and FUV luminosity space was evenly sampled.
The sample was thus not selected for being \lya{} emitters, and consequently show a broad variety of \lya\ properties, ranging from nearly complete absorption (LARS 06, 09) to escape fractions of almost 100\% (LARS 02) \citep{hayes2014}.
The combined sample thus spans a large range both in \lya{} properties and other galaxy properties, such as star formation rate, galaxy mass and morphology and is ideal for use when constructing a multivariate predictive relation.

Each of the 42 galaxies in the sample now has 8 band photometry with the HST (5 broadbands and 3 narrowbands), UV spectroscopy with the \emph{Cosmic Origins Spectrograph}(COS) on the HST, optical SDSS spectra, and 21-cm radio observations with the Very Large Array in D and C-configuration, with further radio observations at higher resolutions currently being performed.

\subsection{Datasets}
	From the raw data a large number of quantities have been derived, many of which have been presented in previous papers \citep{hayes2013,hayes2014,ostlin2014,rivera-thorsen2015}. 
	We subdivide the variables that we consider in our multivariate analysis into two subgroups: direct observables and derived physical quantities. The main reason for this is that using direct observables simplifies application to observational studies and using derived physical quantities enables comparisons with purely theoretical work.

	We later test our results against an additional sample of galaxies observed only with COS, which we describe in more detail in Section~\ref{sec: Out-of-sample}.

	\subsubsection{Observational variables}\label{sec:obs variables}
		
		The \lya{} imaging that is used in this work is derived using a synthesized narrowband approach that was originally described in \citet{hayes2009, hayes2014}. 
		A detailed description of the latest version of the software used, \texttt{LaXs}, will be published in Melinder et al. (in prep). Two of the galaxies (LARS 13, eLARS 12) have net \lya{} absorption and an additional three have signal-to-noise ratios of \lya\ (LARS 10, eLARS 14, eLARS 16) below 1. All of these galaxies are removed from the analysis.

		Our observational dataset consists of 14 variables that are summarized in Table\,\ref{tab: Obs data} and \ref{tab: Obs data 2}. As a quick summary, our dataset consists of 4 broadband filters, 5 nebular lines, one characteristic size measurement and 3 spectrographic properties. Below we give more detailed description of each of the variables.

		\begin{description}
			\item[Broadband luminosities]

			We use broadband luminosities in the far UV (F140LP or F150LP), $U$ (F336W or F390W), $B$ (F438W or F438W), and $I$ (F775W, F850LP or F814W) bands measured in a circular aperture centered on the brightest UV cluster of each galaxy. For the far UV and I band the filter used depends on the redshift of the galaxy, with the highest redshift galaxies (e.g. LARS 14) using the F150LP. The aperture size was defined using the \lya\ image by adding annuli to the aperture until the signal to noise in the next annulus was below 1 or the edge of the image was reached.

			\item[H$\alpha$] The H$\alpha$ luminosity is measured from continuum--subtracted HST narrowband imaging (F673N, F680N, FR656N, FR716N, FR782N depending on galaxy redshift) using the same aperture as the broadband fluxes.

			\item[Other nebular lines] For nebular lines other than H$\alpha$ we used the optical spectrum retrieved from the SDSS.
			The H$\beta$, [OIII] 5007 \AA{}, [OII]3727+3729\AA\ and [NII] 6584\AA\ line luminosities are measured from simultaneously fitting Gaussians, constrained to have the same redshift and velocity width, to the emission lines in the SDSS optical spectrum.
			The SDSS fiber is however considerably smaller than the HST apertures. 
			We therefore scaled the emission lines using the H$\alpha$ fluxes according to 
			$\mathrm{F} = \frac{\mathrm{H}\alpha_{\mathrm{HST}}}{ \mathrm{H}\alpha_{\mathrm{SDSS}}} \cdot \mathrm{F}_{\mathrm{SDSS}}$

			\item[UV size] The size of the galaxies was measured in the far UV, i.e. usually the F140LP HST filter, by using \texttt{GALFIT} \citep{peng2002} to fit a single component 2D Sersić profile to the galaxy. This is, in many cases, not a very good description of the light profile, because a large fraction of the galaxies are complex merger systems. However, a Sersić fit is a simple parametrization that can be used even at high redshifts where spatial resolution is low, which means that using this simple geometric description increases the range of galaxies to which our results can be applied compared to more complex size measurements. Additionally the Sersić formulations was used by \citet{yang2017} to describe their galaxies which means that we can directly use their measurements when testing our predictions (see Section\,\ref{sec: Out-of-sample}).

			\item[v95] The outflow velocity of gas is measured from low ionization (LIS) absorption lines in the COS spectroscopy of the galaxies. The lines used were \siII\,1190, \siII\,1193, \siII\,1260, \siII\,1304, \oI\,1302 and \cII\,1334. The lines were put on a velocity grid centered on the systemic redshift and combined into an average LIS line. v95 is then defined as the velocity that has 95 \% of the absorbed line flux redward of it. More details can be found in \citet{rivera-thorsen2015}

			\item[w90] This measures the width of the averaged LIS lines from the COS spectroscopy and is defined as the velocity width from 5\% to 95\% integrated absorption. More details can be found in \citet{rivera-thorsen2015}

			\item[F$_\mathrm{cov}$] The maximum velocity-binned covering fraction defined as $1-I_\mathrm{min}$ where $I_\mathrm{min}$ is the minimum residual flux in the averaged LIS line. More details can be found in \citet{rivera-thorsen2015}

		\end{description}

\begin{deluxetable*}{lllllllll} 
\tablecaption{Observable variable values, excluding nebular lines, for the LARS (L01 to L14) and eLARS (eL01-eL28) galaxies. \label{tab: Obs data}} 
\tabletypesize{\small}
\tablecolumns{9}
\tablehead{
	\colhead{ID}& 
	\colhead{U \tablenotemark{2}}& 
	\colhead{B \tablenotemark{2}}& 
	\colhead{I \tablenotemark{2}}& 
	\colhead{FUV \tablenotemark{2}}& 
	\colhead{UVsize\tablenotemark{3}}& 
	\colhead{v95\tablenotemark{4}}& 
	\colhead{w90\tablenotemark{4}}& 
	\colhead{Fcov}\\ 
	\colhead{} & 
	\colhead{$10^{40}$} & 
	\colhead{$10^{40}$} & 
	\colhead{$10^{40}$} & 
	\colhead{$10^{40}$} & 
	\colhead{} & 
	\colhead{} & 
	\colhead{} & 
	\colhead{}
}
\startdata
L01 &	$0.538\pm0.0007$ &	$0.4503\pm0.0005$ &	$0.1824\pm0.0002$ &	$1.781\pm0.007$ &	$0.851\pm0.001$ &	$314\pm24$ &	$381\pm24$ &	$0.7\pm0.03$\\ 
L02 &	$0.1853\pm0.0006$ &	$0.1939\pm0.0005$ &	$0.0902\pm0.0002$ &	$0.651\pm0.008$ &	$1.726\pm0.004$ &	$258\pm80$ &	$351\pm102$ &	$0.86\pm0.07$\\ 
L03 &	$0.5582\pm0.0008$ &	$0.7878\pm0.0003$ &	$0.5973\pm0.0002$ &	$0.598\pm0.007$ &	$0.914\pm0.001$ &	$462\pm42$ &	$628\pm51$ &	$0.97\pm0.05$\\ 
L04 &	$0.4054\pm0.0005$ &	$0.378\pm0.0004$ &	$0.1484\pm0.0002$ &	$1.186\pm0.003$ &	$5.909\pm0.026$ &	$261\pm46$ &	$365\pm61$ &	$0.99\pm0.07$\\ 
L05 &	$0.5176\pm0.001$ &	$0.374\pm0.0007$ &	$0.1227\pm0.0003$ &	$2.324\pm0.006$ &	$0.867\pm0.001$ &	$390\pm22$ &	$466\pm47$ &	$0.81\pm0.03$\\ 
L06 &	$0.0189\pm0.0001$ &	$0.0125\pm0.0001$ &	$0.0035\pm0.00002$ &	$0.086\pm0.001$ &	$0.578\pm0.001$ &	$244\pm38$ &	$553\pm43$ &	$0.84\pm0.06$\\ 
L07 &	$0.4709\pm0.0012$ &	$0.4397\pm0.0008$ &	$0.1813\pm0.0004$ &	$1.645\pm0.01$ &	$0.824\pm0.002$ &	$267\pm63$ &	$392\pm68$ &	$0.8\pm0.03$\\ 
L08 &	$1.5318\pm0.0013$ &	$2.0362\pm0.0019$ &	$1.4219\pm0.0006$ &	$2.0\pm0.01$ &	$7.661\pm0.041$ &	$442\pm70$ &	$522\pm85$ &	$0.98\pm0.07$\\ 
L09 &	$2.9033\pm0.0027$ &	$3.2883\pm3.2122$ &	$1.9388\pm0.0016$ &	$5.004\pm0.022$ &	$21.238\pm0.112$ &	$263\pm27$ &	$503\pm34$ &	$1.0\pm0.05$\\ 
L10 &	$0.5508\pm0.0016$ &	$0.6839\pm0.0011$ &	$0.3905\pm0.0005$ &	$1.118\pm0.013$ &	$2.659\pm0.003$ &	$287\pm38$ &	$484\pm51$ &	$1.03\pm0.08$\\ 
L11 &	$3.8293\pm0.0083$ &	$4.3231\pm0.0058$ &	$2.8447\pm0.0028$ &	$8.508\pm0.074$ &	$19.641\pm0.043$ &	$398\pm68$ &	$433\pm176$ &	$1.04\pm0.09$\\ 
L12 &	$2.0826\pm0.0086$ &	$1.6668\pm0.0065$ &	$0.7601\pm0.0031$ &	$7.849\pm0.086$ &	$0.881\pm0.001$ &	$289\pm54$ &	$503\pm56$ &	$0.82\pm0.03$\\ 
L13 &	$2.3499\pm0.0091$ &	$2.1426\pm0.0116$ &	$1.0142\pm0.0106$ &	$5.891\pm0.175$ &	$0.964\pm0.001$ &	$359\pm36$ &	$484\pm43$ &	$0.76\pm0.04$\\ 
L14 &	$1.9181\pm0.0156$ &	$1.4906\pm0.0097$ &	$0.9119\pm0.0044$ &	$10.638\pm0.352$ &	$0.708\pm0.0004$ &	$461\pm110$ &	$485\pm122$ &	$0.4\pm0.05$\\ 
eL01 &	$1.6182\pm0.0013$ &	$1.7599\pm0.0003$ &	$1.1892\pm0.0002$ &	$2.322\pm0.003$ &	$1.003\pm0.001$ &	$656\pm15$ &	$658\pm15$ &	$0.89\pm0.01$\\ 
eL02 &	$1.0836\pm0.0017$ &	$1.1768\pm0.0011$ &	$0.5588\pm0.0008$ &	$2.636\pm0.007$ &	$3.335\pm0.003$ &	$337\pm87$ &	$346\pm87$ &	$0.92\pm0.04$\\ 
eL03 &	$0.0997\pm0.0005$ &	$0.0998\pm0.0003$ &	$0.047\pm0.0002$ &	$0.228\pm0.004$ &	$18.091\pm0.179$ &	$370\pm109$ &	$538\pm95$ &	$0.94\pm0.07$\\ 
eL04 &	$1.0005\pm0.0014$ &	$1.1619\pm0.001$ &	$0.566\pm0.0004$ &	$2.063\pm0.093$ &	$2.984\pm0.003$ &	$560\pm51$ &	$600\pm59$ &	$0.89\pm0.04$\\ 
eL05 &	$1.0924\pm0.0018$ &	$1.5318\pm0.0015$ &	$1.0544\pm0.0006$ &	$2.274\pm0.013$ &	$22.59\pm0.118$ &	$528\pm97$ &	$553\pm102$ &	$0.62\pm0.03$\\ 
eL06 &	$0.4253\pm0.0015$ &	$0.4874\pm0.0009$ &	$0.2444\pm0.0004$ &	$0.929\pm0.011$ &	$4.166\pm0.005$ &	$227\pm160$ &	$277\pm160$ &	$0.84\pm0.14$\\ 
eL07 &	$0.3037\pm0.0017$ &	$0.2804\pm0.001$ &	$0.1266\pm0.0006$ &	$1.05\pm0.013$ &	$3.485\pm0.011$ &	$215\pm126$ &	$349\pm124$ &	$0.61\pm0.06$\\ 
eL08 &	$0.5642\pm0.0013$ &	$0.8436\pm0.0008$ &	$0.5894\pm0.0004$ &	$0.818\pm0.009$ &	$18.329\pm0.066$ &	$383\pm117$ &	$482\pm117$ &	$0.94\pm0.07$\\ 
eL09 &	$0.2963\pm0.0008$ &	$0.3682\pm0.0006$ &	$0.1568\pm0.0003$ &	$0.73\pm0.006$ &	$1.167\pm0.001$ &	$420\pm51$ &	$526\pm44$ &	$0.85\pm0.02$\\ 
eL10 &	$0.3365\pm0.0013$ &	$0.484\pm0.0008$ &	$0.3228\pm0.0004$ &	$0.55\pm0.034$ &	$11.619\pm0.036$ &	$169\pm124$ &	$306\pm148$ &	$0.91\pm0.33$\\ 
eL11 &	$0.2749\pm0.0008$ &	$0.3621\pm0.0005$ &	$0.2189\pm0.0003$ &	$0.554\pm0.005$ &	$2.215\pm0.002$ &	$249\pm132$ &	$395\pm132$ &	$0.95\pm0.17$\\ 
eL12 &	$0.4662\pm0.0009$ &	$0.668\pm0.0006$ &	$0.4658\pm0.0003$ &	$0.578\pm0.006$ &	$5.314\pm0.002$ &	$427\pm204$ &	$438\pm233$ &	$1.0\pm0.68$\\ 
eL13 &	$0.2347\pm0.0006$ &	$0.1907\pm0.0004$ &	$0.0999\pm0.0002$ &	$0.659\pm0.004$ &	$0.137\pm0.00006$ &	$339\pm22$ &	$365\pm87$ &	$0.45\pm0.01$\\ 
eL14 &	$0.2243\pm0.0007$ &	$0.2577\pm0.0004$ &	$0.1264\pm0.0002$ &	$0.476\pm0.005$ &	$1.559\pm0.001$ &	$129\pm87$ &	$248\pm95$ &	$0.85\pm0.1$\\ 
eL15 &	$0.2637\pm0.0014$ &	$0.419\pm0.001$ &	$0.261\pm0.0004$ &	$0.374\pm0.009$ &	$1.684\pm0.002$ &	$274\pm102$ &	$334\pm109$ &	$0.84\pm0.12$\\ 
eL16 &	$0.12\pm0.0007$ &	$0.1838\pm0.0004$ &	$0.1098\pm0.0002$ &	$0.238\pm0.004$ &	$16.337\pm0.12$ &	$258\pm175$ &	$422\pm189$ &	$0.8\pm0.18$\\ 
eL17 &	$0.2093\pm0.0011$ &	$0.3029\pm0.0007$ &	$0.1932\pm0.0004$ &	$0.415\pm0.008$ &	$8.316\pm0.042$ &	$136\pm153$ &	$263\pm161$ &	$0.8\pm0.22$\\ 
eL18 &	$0.0899\pm0.0008$ &	$0.1182\pm0.0007$ &	$0.068\pm0.0002$ &	$0.218\pm0.006$ &	$4.311\pm0.005$ &	$239\pm117$ &	$366\pm132$ &	$0.73\pm0.16$\\ 
eL19 &	$0.0856\pm0.0006$ &	$0.0931\pm0.0004$ &	$0.0397\pm0.0002$ &	$0.24\pm0.004$ &	$1.792\pm0.002$ &	$210\pm175$ &	$365\pm183$ &	$0.85\pm0.06$\\ 
eL20 &	$0.1476\pm0.0006$ &	$0.2007\pm0.0004$ &	$0.1421\pm0.0002$ &	$0.241\pm0.004$ &	$1.866\pm0.002$ &	$264\pm102$ &	$350\pm109$ &	$0.79\pm0.09$\\ 
eL21 &	$0.0533\pm0.0005$ &	$0.0654\pm0.0003$ &	$0.0334\pm0.0002$ &	$0.116\pm0.003$ &	$5.636\pm0.026$ &	$199\pm141$ &	$277\pm153$ &	$0.52\pm0.22$\\ 
eL22 &	$0.8044\pm0.0025$ &	$0.7607\pm0.0016$ &	$0.3424\pm0.0011$ &	$2.284\pm0.016$ &	$2.906\pm0.007$ &	$341\pm57$ &	$474\pm57$ &	$0.85\pm0.01$\\ 
eL23 &	$1.239\pm0.003$ &	$1.6177\pm0.0019$ &	$0.9917\pm0.0015$ &	$2.043\pm0.021$ &	$30.257\pm0.42$ &	$319\pm122$ &	$415\pm136$ &	$0.88\pm0.18$\\ 
eL24 &	$1.1523\pm0.0024$ &	$1.5559\pm0.0019$ &	$1.2807\pm0.0013$ &	$1.858\pm0.017$ &	$1.937\pm0.002$ &	$612\pm158$ &	$862\pm158$ &	$0.7\pm0.04$\\ 
eL25 &	$0.6047\pm0.0027$ &	$0.7293\pm0.0017$ &	$0.4303\pm0.0013$ &	$1.568\pm0.019$ &	$13.786\pm0.065$ &	$347\pm118$ &	$447\pm122$ &	$0.77\pm0.12$\\ 
eL26 &	$0.6352\pm0.0026$ &	$0.8996\pm0.0017$ &	$0.6117\pm0.0013$ &	$1.047\pm0.018$ &	$10.222\pm0.045$ &	$255\pm86$ &	$317\pm101$ &	$0.85\pm0.14$\\ 
eL27 &	$0.4971\pm0.002$ &	$0.6126\pm0.0013$ &	$0.3297\pm0.0009$ &	$1.049\pm0.012$ &	$6.686\pm0.014$ &	$255\pm180$ &	$317\pm223$ &	$0.69\pm0.46$\\ 
eL28 &	$0.488\pm0.0018$ &	$0.5787\pm0.0012$ &	$0.304\pm0.0008$ &	$0.969\pm0.012$ &	$8.935\pm0.022$ &	$293\pm43$ &	$317\pm86$ &	$0.83\pm0.07$\\  
\enddata
		\tablenotetext{1}{[erg s$^{-1}$]}
		\tablenotetext{2}{[erg s$^{-1}$\AA$^{-1}$]}
		\tablenotetext{3}{[kpc]}
		\tablenotetext{4}{[km s$^{-1}$]}
\end{deluxetable*}

\begin{deluxetable*}{lllllll} 
\tablecaption{Nebular line measurements for the data set of observable variables\label{tab: Obs data 2}} 
\tablecolumns{7}
\tablehead{
	\colhead{ID}& 
	\colhead{Ly$\alpha$ \tablenotemark{1}}& 
	\colhead{H$\alpha$\tablenotemark{1}}& 
	\colhead{H$\beta$\tablenotemark{1}}& 
	\colhead{[OIII]\tablenotemark{1}}& 
	\colhead{[OII]\tablenotemark{1}}& 
	\colhead{[NII]\tablenotemark{1}}\\ 
	\colhead{} & 
	\colhead{$10^{41}$} & 
	\colhead{$10^{40}$} & 
	\colhead{$10^{40}$} & 
	\colhead{$10^{40}$} & 
	\colhead{$10^{40}$} & 
	\colhead{$10^{40}$}
}
\startdata
L01 &	$8.91\pm0.09$ &	$51.21\pm0.11$ &	$16.1\pm0.05$ &	$68.92\pm0.21$ &	$16.72\pm0.05$ &	$2.75\pm0.01$\\ 
L02 &	$4.27\pm0.08$ &	$15.9\pm0.11$ &	$5.18\pm0.07$ &	$23.43\pm0.34$ &	$5.15\pm0.07$ &	$0.9\pm0.02$\\ 
L03 &	$1.73\pm0.08$ &	$60.7\pm0.23$ &	$10.73\pm0.05$ &	$11.49\pm0.05$ &	$8.59\pm0.04$ &	$23.7\pm0.11$\\ 
L04 &	$0.52\pm0.05$ &	$38.99\pm0.09$ &	$10.85\pm0.03$ &	$52.74\pm0.16$ &	$12.07\pm0.04$ &	$2.06\pm0.01$\\ 
L05 &	$7.14\pm0.11$ &	$42.6\pm0.18$ &	$12.96\pm0.07$ &	$65.75\pm0.36$ &	$9.99\pm0.05$ &	$1.67\pm0.01$\\ 
L06 &	$0.02\pm0.01$ &	$2.37\pm0.01$ &	$0.79\pm0.00002$ &	$3.71\pm0.01$ &	$0.74\pm0.0$ &	$0.08\pm0.01$\\ 
L07 &	$6.9\pm0.13$ &	$37.91\pm0.14$ &	$10.97\pm0.04$ &	$45.85\pm0.15$ &	$10.49\pm0.03$ &	$3.33\pm0.01$\\ 
L08 &	$3.94\pm0.12$ &	$114.21\pm0.51$ &	$27.84\pm1.99$ &	$57.49\pm4.11$ &	$31.2\pm2.23$ &	$28.97\pm2.07$\\ 
L09 &	$6.51\pm0.27$ &	$189.59\pm0.3$ &	$48.62\pm0.16$ &	$215.91\pm0.7$ &	$36.53\pm0.12$ &	$20.47\pm0.07$\\ 
L10 &	$0.11\pm0.12$ &	$20.02\pm0.2$ &	$5.29\pm0.06$ &	$9.82\pm0.11$ &	$6.65\pm0.07$ &	$3.45\pm0.04$\\ 
L11 &	$18.17\pm0.62$ &	$99.75\pm0.66$ &	$22.26\pm0.32$ &	$19.88\pm0.29$ &	$23.02\pm0.33$ &	$30.02\pm0.44$\\ 
L12 &	$17.0\pm0.62$ &	$139.12\pm1.15$ &	$41.24\pm0.26$ &	$183.81\pm1.14$ &	$32.65\pm0.2$ &	$11.44\pm0.08$\\ 
L13 &	$-19.7\pm4.2$ &	$210.03\pm1.17$ &	$53.35\pm0.42$ &	$132.59\pm1.02$ &	$50.45\pm0.39$ &	$43.32\pm0.54$\\ 
L14 &	$56.66\pm2.51$ &	$189.92\pm2.45$ &	$58.47\pm0.56$ &	$323.95\pm3.1$ &	$40.02\pm0.38$ &	$6.37\pm0.72$\\ 
eL01 &	$5.5\pm0.08$ &	$131.25\pm0.15$ &	$28.36\pm0.57$ &	$12.91\pm0.28$ &	$20.85\pm0.44$ &	$54.48\pm1.05$\\ 
eL02 &	$3.56\pm0.09$ &	$61.69\pm0.28$ &	$13.34\pm0.61$ &	$22.95\pm1.04$ &	$16.57\pm0.78$ &	$15.03\pm0.67$\\ 
eL03 &	$0.08\pm0.05$ &	$6.22\pm0.07$ &	$1.18\pm0.02$ &	$0.82\pm0.02$ &	$1.17\pm0.04$ &	$2.1\pm0.03$\\ 
eL04 &	$4.71\pm0.14$ &	$42.16\pm0.17$ &	$10.24\pm0.3$ &	$12.54\pm0.37$ &	$15.35\pm0.48$ &	$10.3\pm0.3$\\ 
eL05 &	$5.82\pm0.18$ &	$36.61\pm0.19$ &	$7.69\pm0.29$ &	$15.14\pm0.54$ &	$11.19\pm0.55$ &	$19.39\pm0.69$\\ 
eL06 &	$1.38\pm0.12$ &	$13.12\pm0.18$ &	$3.15\pm0.27$ &	$2.78\pm0.24$ &	$4.27\pm0.38$ &	$3.44\pm0.29$\\ 
eL07 &	$1.12\pm0.15$ &	$25.52\pm0.29$ &	$7.22\pm0.26$ &	$38.05\pm1.34$ &	$6.61\pm0.27$ &	$0.87\pm0.04$\\ 
eL08 &	$1.26\pm0.1$ &	$17.35\pm0.18$ &	$3.4\pm0.22$ &	$0.92\pm0.08$ &	$2.67\pm0.21$ &	$6.52\pm0.4$\\ 
eL09 &	$0.45\pm0.07$ &	$7.53\pm0.12$ &	$2.22\pm0.06$ &	$8.79\pm0.22$ &	$3.76\pm0.11$ &	$0.57\pm0.02$\\ 
eL10 &	$0.98\pm0.12$ &	$13.82\pm0.2$ &	$2.49\pm0.12$ &	$1.71\pm0.08$ &	$3.42\pm0.17$ &	$4.59\pm0.19$\\ 
eL11 &	$0.73\pm0.1$ &	$8.99\pm0.17$ &	$2.38\pm0.11$ &	$4.71\pm0.21$ &	$4.17\pm0.2$ &	$2.57\pm0.12$\\ 
eL12 &	$-0.28\pm0.08$ &	$22.55\pm0.15$ &	$4.02\pm0.22$ &	$1.98\pm0.12$ &	$5.18\pm0.31$ &	$7.87\pm0.42$\\ 
eL13 &	$2.62\pm0.08$ &	$8.83\pm0.07$ &	$2.28\pm0.03$ &	$2.37\pm0.03$ &	$1.69\pm0.03$ &	$2.41\pm0.03$\\ 
eL14 &	$0.05\pm0.06$ &	$10.78\pm0.1$ &	$2.67\pm0.08$ &	$5.83\pm0.16$ &	$4.01\pm0.13$ &	$1.78\pm0.05$\\ 
eL15 &	$0.74\pm0.1$ &	$6.41\pm0.18$ &	$1.44\pm0.1$ &	$2.41\pm0.16$ &	$2.24\pm0.17$ &	$1.68\pm0.11$\\ 
eL16 &	$0.03\pm0.05$ &	$3.4\pm0.1$ &	$0.86\pm0.05$ &	$1.37\pm0.07$ &	$1.36\pm0.08$ &	$0.61\pm0.03$\\ 
eL17 &	$0.77\pm0.09$ &	$5.76\pm0.14$ &	$1.34\pm0.11$ &	$0.97\pm0.09$ &	$1.38\pm0.13$ &	$1.69\pm0.14$\\ 
eL18 &	$0.2\pm0.06$ &	$3.54\pm0.14$ &	$0.96\pm0.13$ &	$2.55\pm0.34$ &	$1.52\pm0.21$ &	$0.34\pm0.05$\\ 
eL19 &	$0.32\pm0.05$ &	$4.3\pm0.09$ &	$1.12\pm0.04$ &	$4.54\pm0.16$ &	$1.2\pm0.05$ &	$0.23\pm0.01$\\ 
eL20 &	$0.23\pm0.04$ &	$5.19\pm0.08$ &	$1.35\pm0.05$ &	$0.92\pm0.04$ &	$1.64\pm0.07$ &	$1.34\pm0.05$\\ 
eL21 &	$0.12\pm0.04$ &	$1.19\pm0.1$ &	$0.33\pm0.04$ &	$1.06\pm0.12$ &	$0.46\pm0.05$ &	$0.08\pm0.01$\\ 
eL22 &	$2.2\pm0.16$ &	$36.14\pm0.53$ &	$9.91\pm0.37$ &	$29.69\pm1.09$ &	$14.61\pm0.57$ &	$3.04\pm0.12$\\ 
eL23 &	$1.21\pm0.24$ &	$42.83\pm0.73$ &	$8.43\pm0.91$ &	$5.09\pm0.56$ &	$9.94\pm1.09$ &	$13.92\pm1.49$\\ 
eL24 &	$4.81\pm0.16$ &	$67.62\pm0.62$ &	$15.37\pm0.29$ &	$21.47\pm0.4$ &	$23.49\pm0.5$ &	$24.75\pm0.45$\\ 
eL25 &	$1.87\pm0.22$ &	$19.42\pm0.52$ &	$5.1\pm0.7$ &	$4.77\pm0.66$ &	$7.08\pm0.98$ &	$4.32\pm0.6$\\ 
eL26 &	$2.14\pm0.16$ &	$32.13\pm0.53$ &	$5.49\pm0.78$ &	$2.19\pm0.33$ &	$4.19\pm0.62$ &	$9.63\pm1.36$\\ 
eL27 &	$2.06\pm0.16$ &	$18.54\pm0.36$ &	$5.31\pm0.85$ &	$4.71\pm0.76$ &	$7.43\pm1.2$ &	$4.45\pm0.71$\\ 
eL28 &	$0.43\pm0.12$ &	$23.02\pm0.36$ &	$5.85\pm0.28$ &	$8.19\pm0.38$ &	$7.59\pm0.39$ &	$4.36\pm0.21$\\ 
\enddata
		\tablenotetext{1}{[erg s$^{-1}$]}
\end{deluxetable*}

	\subsubsection{Derived physical quantities}
		From the observational data we derived some commonly used physical properties of the galaxies that we present in this section. 

		\begin{description}
			\item[M$_{*}$]
				Stellar mass --- derived from pixel--wise SED modeling of the HST broadband observations in the same aperture as used for broadband luminosities in section \ref{sec:obs variables}.
			\item[0$_{32}$]
				Since the true ionization parameter cannot be determined, we use the standard observational diagnostic defined as the ratio of [OIII] 5007 / [OII] 3727+3729 which correlates strongly with the ionization parameter \citep{kewley2002,kobulnicky2004}
			\item[E$_{(B-V)}$]
				Dust extinction derived from the H$\alpha$ / H$\beta$ ratio in the SDSS spectrum using the CCM extinction law \citep{cardelli1989}.
			\item[SFR] 
				Star formation rate derived from H$\alpha$ using the Kennicut-calibration \citep{kennicuttjr.1998}. The H$\alpha$ flux is measured in the HST imaging, but due to the large uncertainties in the H$\beta$ image we use the E$_{\mathrm{B}-\mathrm{V}}$ derived from the SDSS spectroscopy and the CCM law \citep{cardelli1989} for the dust correction.
			\item[log(O/H) + 12]
				Nebular oxygen abundance derived from the O3N2 strong line calibration \citep{yin2007}

		\end{description}

		In addition to these quantities we also include UV size, $v95$ and $w90$ as described in the previous section since these can be considered both directly observables and physical properties of the systems. The data used are shown in Table \ref{tab: Phys data}

        \begin{deluxetable*}{llllllllll}[htb!]
          \tabletypesize{\small}     
          \tablecaption{Physical variables derived for the lars (l01 to l14) and elars (el01-el28) galaxies. \label{tab: Phys data}}
          \tablecolumns{10}
\tablehead{\colhead{ID}& 
	\colhead{M$_*$\tablenotemark{1}}& 
	\colhead{UVsize\tablenotemark{2}}& 
	\colhead{E$_{B-V}$\tablenotemark{3}}& 
	\colhead{O32}& 
	\colhead{O/H}& 
	\colhead{SFR\tablenotemark{4}}& 
	\colhead{v95\tablenotemark{5}}& 
	\colhead{w90\tablenotemark{5}}& 
	\colhead{F$_C$}\\
	\colhead{}& 
	\colhead{$10^{10}$}& 
	\colhead{}& 
	\colhead{}& 
	\colhead{}& 
	\colhead{}& 
	\colhead{}& 
	\colhead{}& 
	\colhead{}& 
	\colhead{}
}
\startdata
L01 &	$1.62\pm0.04$ &	$0.8511\pm0.0012$ &	$0.1076\pm0.0001$ &	$4.1215\pm0.0006$ &	$8.2182\pm0.0001$ &	$3.92\pm0.01$ &	$314\pm24$ &	$381\pm24$ &	$0.7\pm0.03$\\ 
L02 &	$0.58\pm0.03$ &	$1.7256\pm0.0041$ &	$0.0716\pm0.0002$ &	$4.5481\pm0.0004$ &	$8.2186\pm0.004$ &	$1.09\pm0.01$ &	$258\pm80$ &	$351\pm102$ &	$0.86\pm0.07$\\ 
L03 &	$2.28\pm0.45$ &	$0.9136\pm0.0014$ &	$0.6897\pm0.0001$ &	$1.3374\pm0.0001$ &	$8.4163\pm0.00001$ &	$27.61\pm0.1$ &	$462\pm42$ &	$628\pm51$ &	$0.97\pm0.05$\\ 
L04 &	$1.11\pm0.01$ &	$5.9088\pm0.026$ &	$0.2312\pm0.00003$ &	$4.3706\pm0.0002$ &	$8.1787\pm0.0003$ &	$4.35\pm0.01$ &	$261\pm46$ &	$365\pm61$ &	$0.99\pm0.07$\\ 
L05 &	$0.86\pm0.03$ &	$0.8669\pm0.0007$ &	$0.1407\pm0.00004$ &	$6.5808\pm0.0008$ &	$8.0747\pm0.0004$ &	$3.6\pm0.02$ &	$390\pm22$ &	$466\pm47$ &	$0.81\pm0.03$\\ 
L06 &	$0.02\pm0.0$ &	$0.5782\pm0.0009$ &	$0.0485\pm0.0006$ &	$4.9972\pm0.0009$ &	$8.0638\pm0.0351$ &	$0.15\pm0.0$ &	$244\pm38$ &	$553\pm43$ &	$0.84\pm0.06$\\ 
L07 &	$0.9\pm0.03$ &	$0.8244\pm0.0017$ &	$0.1911\pm0.00004$ &	$4.3694\pm0.0003$ &	$8.3396\pm0.0001$ &	$3.74\pm0.01$ &	$267\pm63$ &	$392\pm68$ &	$0.8\pm0.03$\\ 
L08 &	$10.35\pm0.47$ &	$7.6614\pm0.0413$ &	$0.3645\pm0.0009$ &	$1.8424\pm0.0009$ &	$8.5055\pm0.00003$ &	$19.19\pm0.1$ &	$442\pm70$ &	$522\pm85$ &	$0.98\pm0.07$\\ 
L09 &	$5.84\pm0.07$ &	$21.2382\pm0.1115$ &	$0.3134\pm0.00003$ &	$5.9102\pm0.0003$ &	$8.3679\pm0.00003$ &	$27.23\pm0.04$ &	$263\pm27$ &	$503\pm34$ &	$1.0\pm0.05$\\ 
L10 &	$1.81\pm0.04$ &	$2.6594\pm0.0031$ &	$0.2823\pm0.0006$ &	$1.4772\pm0.0006$ &	$8.5049\pm0.0001$ &	$2.61\pm0.03$ &	$287\pm38$ &	$484\pm51$ &	$1.03\pm0.08$\\ 
L11 &	$15.06\pm0.32$ &	$19.6407\pm0.0429$ &	$0.4538\pm0.0014$ &	$0.8637\pm0.0017$ &	$8.4277\pm0.0004$ &	$22.02\pm0.18$ &	$398\pm68$ &	$433\pm176$ &	$1.04\pm0.09$\\ 
L12 &	$4.28\pm0.42$ &	$0.8814\pm0.0008$ &	$0.1669\pm0.0001$ &	$5.6296\pm0.0004$ &	$8.3121\pm0.0006$ &	$12.76\pm0.11$ &	$289\pm54$ &	$503\pm56$ &	$0.82\pm0.03$\\ 
L13 &	$3.73\pm0.66$ &	$0.9639\pm0.0011$ &	$0.3229\pm0.0019$ &	$2.6281\pm0.0032$ &	$8.5019\pm0.0003$ &	$31.06\pm0.26$ &	$359\pm36$ &	$484\pm43$ &	$0.76\pm0.04$\\ 
L14 &	$3.83\pm0.96$ &	$0.7083\pm0.0004$ &	$0.1287\pm0.0007$ &	$8.0942\pm0.0039$ &	$7.9919\pm0.0416$ &	$15.49\pm0.21$ &	$461\pm110$ &	$485\pm122$ &	$0.4\pm0.05$\\ 
EL01 &	$7.75\pm0.05$ &	$1.0035\pm0.0011$ &	$0.4867\pm0.0139$ &	$0.6193\pm0.0114$ &	$8.2278\pm0.0054$ &	$32.05\pm1.38$ &	$656\pm15$ &	$658\pm15$ &	$0.89\pm0.01$\\ 
EL02 &	$2.98\pm0.02$ &	$3.3352\pm0.0026$ &	$0.4853\pm0.0192$ &	$1.3851\pm0.0371$ &	$8.5022\pm0.0009$ &	$15.0\pm0.9$ &	$337\pm87$ &	$346\pm87$ &	$0.92\pm0.04$\\ 
EL03 &	$0.19\pm0.01$ &	$18.0912\pm0.1795$ &	$0.616\pm0.0174$ &	$0.695\pm0.0254$ &	$8.367\pm0.0056$ &	$2.26\pm0.13$ &	$370\pm109$ &	$538\pm95$ &	$0.94\pm0.07$\\ 
EL04 &	$3.33\pm0.03$ &	$2.9836\pm0.0033$ &	$0.3686\pm0.0138$ &	$0.817\pm0.0144$ &	$8.4838\pm0.0015$ &	$7.17\pm0.31$ &	$560\pm51$ &	$600\pm59$ &	$0.89\pm0.04$\\ 
EL05 &	$6.91\pm0.08$ &	$22.5895\pm0.118$ &	$0.5158\pm0.0217$ &	$1.3531\pm0.052$ &	$8.456\pm0.0031$ &	$9.77\pm0.66$ &	$528\pm97$ &	$553\pm102$ &	$0.62\pm0.03$\\ 
EL06 &	$1.47\pm0.04$ &	$4.1655\pm0.0053$ &	$0.3803\pm0.025$ &	$0.6497\pm0.0255$ &	$8.4442\pm0.0047$ &	$2.31\pm0.18$ &	$227\pm160$ &	$277\pm160$ &	$0.84\pm0.14$\\ 
EL07 &	$1.69\pm0.1$ &	$3.4854\pm0.0115$ &	$0.2137\pm0.0176$ &	$5.7574\pm0.1313$ &	$8.0152\pm0.0103$ &	$2.7\pm0.15$ &	$215\pm126$ &	$349\pm124$ &	$0.61\pm0.06$\\ 
EL08 &	$3.83\pm0.07$ &	$18.329\pm0.0661$ &	$0.5851\pm0.0252$ &	$0.3456\pm0.0266$ &	$8.1079\pm0.0195$ &	$5.73\pm0.46$ &	$383\pm117$ &	$482\pm117$ &	$0.94\pm0.07$\\ 
EL09 &	$0.47\pm0.03$ &	$1.167\pm0.0008$ &	$0.1743\pm0.0198$ &	$2.341\pm0.0458$ &	$8.3176\pm0.0069$ &	$0.71\pm0.05$ &	$420\pm51$ &	$526\pm44$ &	$0.85\pm0.02$\\ 
EL10 &	$1.88\pm0.07$ &	$11.6191\pm0.0357$ &	$0.6713\pm0.0266$ &	$0.4988\pm0.0201$ &	$8.3688\pm0.0071$ &	$5.94\pm0.5$ &	$169\pm124$ &	$306\pm148$ &	$0.91\pm0.33$\\ 
EL11 &	$2.54\pm0.1$ &	$2.2148\pm0.0022$ &	$0.2801\pm0.0193$ &	$1.1299\pm0.0262$ &	$8.5014\pm0.001$ &	$1.17\pm0.07$ &	$249\pm132$ &	$395\pm132$ &	$0.95\pm0.17$\\ 
EL12 &	$2.7\pm0.08$ &	$5.3142\pm0.0023$ &	$0.6803\pm0.0217$ &	$0.3831\pm0.0171$ &	$8.2901\pm0.0091$ &	$9.97\pm0.68$ &	$427\pm204$ &	$438\pm233$ &	$1.0\pm0.66$\\ 
EL13 &	$0.68\pm0.01$ &	$0.1368\pm0.0001$ &	$0.3069\pm0.0107$ &	$1.4057\pm0.0265$ &	$8.4589\pm0.0016$ &	$1.24\pm0.04$ &	$339\pm22$ &	$365\pm87$ &	$0.45\pm0.01$\\ 
EL14 &	$0.59\pm0.04$ &	$1.5586\pm0.0008$ &	$0.3484\pm0.0173$ &	$1.4566\pm0.034$ &	$8.4981\pm0.0012$ &	$1.72\pm0.09$ &	$129\pm87$ &	$248\pm95$ &	$0.85\pm0.1$\\ 
EL15 &	$1.09\pm0.04$ &	$1.6836\pm0.0021$ &	$0.4497\pm0.0256$ &	$1.0751\pm0.0416$ &	$8.4984\pm0.0017$ &	$1.4\pm0.12$ &	$274\pm102$ &	$334\pm109$ &	$0.84\pm0.12$\\ 
EL16 &	$0.62\pm0.02$ &	$16.3366\pm0.1196$ &	$0.3221\pm0.0279$ &	$1.013\pm0.0398$ &	$8.5064\pm0.0002$ &	$0.5\pm0.05$ &	$258\pm175$ &	$422\pm189$ &	$0.8\pm0.18$\\ 
EL17 &	$1.29\pm0.04$ &	$8.316\pm0.0416$ &	$0.4122\pm0.0281$ &	$0.7062\pm0.0369$ &	$8.4\pm0.0075$ &	$1.12\pm0.1$ &	$136\pm153$ &	$263\pm161$ &	$0.8\pm0.21$\\ 
EL18 &	$0.54\pm0.2$ &	$4.3111\pm0.0047$ &	$0.259\pm0.033$ &	$1.6726\pm0.0727$ &	$8.43\pm0.0076$ &	$0.43\pm0.05$ &	$239\pm117$ &	$366\pm132$ &	$0.73\pm0.16$\\ 
EL19 &	$0.33\pm0.02$ &	$1.7917\pm0.0016$ &	$0.3025\pm0.0158$ &	$3.7944\pm0.0864$ &	$8.2358\pm0.0077$ &	$0.6\pm0.03$ &	$210\pm175$ &	$365\pm183$ &	$0.85\pm0.06$\\ 
EL20 &	$4.03\pm0.47$ &	$1.8663\pm0.0017$ &	$0.2995\pm0.0204$ &	$0.5578\pm0.0186$ &	$8.4096\pm0.0052$ &	$0.71\pm0.05$ &	$264\pm102$ &	$350\pm109$ &	$0.79\pm0.09$\\ 
EL21 &	$0.21\pm0.01$ &	$5.6357\pm0.0261$ &	$0.2348\pm0.0325$ &	$2.2992\pm0.1015$ &	$8.3374\pm0.0159$ &	$0.13\pm0.02$ &	$199\pm141$ &	$277\pm153$ &	$0.52\pm0.22$\\ 
EL22 &	$0.97\pm0.14$ &	$2.9055\pm0.0071$ &	$0.246\pm0.0156$ &	$2.0325\pm0.0376$ &	$8.393\pm0.0042$ &	$4.22\pm0.22$ &	$341\pm57$ &	$474\pm57$ &	$0.85\pm0.01$\\ 
EL23 &	$5.01\pm0.1$ &	$30.2572\pm0.4199$ &	$0.5807\pm0.0258$ &	$0.5118\pm0.021$ &	$8.3488\pm0.0086$ &	$13.95\pm1.16$ &	$319\pm122$ &	$415\pm136$ &	$0.88\pm0.18$\\ 
EL24 &	$8.52\pm0.12$ &	$1.9373\pm0.0021$ &	$0.435\pm0.0117$ &	$0.9139\pm0.0146$ &	$8.4588\pm0.0018$ &	$14.1\pm0.53$ &	$612\pm158$ &	$862\pm158$ &	$0.7\pm0.04$\\ 
EL25 &	$3.46\pm0.2$ &	$13.7863\pm0.065$ &	$0.2892\pm0.0243$ &	$0.6726\pm0.023$ &	$8.4685\pm0.0038$ &	$2.59\pm0.21$ &	$347\pm118$ &	$447\pm122$ &	$0.77\pm0.13$\\ 
EL26 &	$4.48\pm0.08$ &	$10.2224\pm0.0454$ &	$0.7244\pm0.0243$ &	$0.5232\pm0.0358$ &	$8.2764\pm0.0132$ &	$16.25\pm1.27$ &	$255\pm86$ &	$317\pm101$ &	$0.85\pm0.14$\\ 
EL27 &	$1.93\pm0.03$ &	$6.6857\pm0.0138$ &	$0.202\pm0.0285$ &	$0.6335\pm0.0266$ &	$8.4553\pm0.0051$ &	$1.89\pm0.17$ &	$255\pm180$ &	$317\pm223$ &	$0.69\pm0.48$\\ 
EL28 &	$1.23\pm0.05$ &	$8.9355\pm0.022$ &	$0.323\pm0.0197$ &	$1.0799\pm0.031$ &	$8.5035\pm0.0008$ &	$3.41\pm0.22$ &	$293\pm43$ &	$317\pm86$ &	$0.83\pm0.07$\\ 
\enddata
\tablenotetext{1}{[M$_\odot$]}
\tablenotetext{2}{[kpc]}
\tablenotetext{3}{[mag]}
\tablenotetext{4}{[M$_\odot$ yr$^{-1}$]}
\tablenotetext{5}{[km s$^{-1}$]}
    	\end{deluxetable*}


\section{Methods}\label{sec: Methods}

\subsection{Fitting method considerations}
	As mentioned in the introduction, the main aim of the current work is to address the question of \lya\ production and escape from a multivariate standpoint and use a large dataset to produce a relation that can accurately predict the global \lya\ luminosity. There are many different methods of varying complexity that can be used to accomplish this, ranging from simple regression techniques to neural networks. 
	While more complex machine learning techniques have become increasingly popular in astronomy due to their predictive power when applied to large datasets, that power comes at a cost. 
	In this case the cost is twofold: overfitting and lack of interpretability.

	The most important issue is overfitting and it must be carefully considered when there is a large number of variables in the dataset and a comparatively small number of galaxies.
	A prediction method that has a large number of tunable parameters, such as neural network, will inevitably become very tailored to the specifics of the dataset.
	Such a model would give predictions that are very accurate for our particular galaxies but most likely wildly inaccurate for galaxies outside our sample, which would defeat the purpose of the current work (see section \ref{sec: Out-of-sample} for more details).

	The interpretability or ``black box'' problem is a general issue common to most types of advanced machine learning (see for instance \citealt{morice-atkinson2018}.)
	The issue arises from the fact that when the model becomes sufficiently complex it becomes hard or even impossible to show how the computer arrived at a given set of predictions, due to the large number of adjustable parameters and the nonlinearity of the weight assignment process.
	If the goal is for the relation to be simple to use by others and also that it should provide some additional understanding of the underlying physical processes, the model must be kept simple.

	Established techniques that have been used in other multivariate problems, such as Principal Component Analysis, are not particularly well-suited to the analysis we wish to conduct for two reasons.
	The first is that PCA searches for the direction of minimal overall dispersion within a dataset with no variable being given any special importance, but we instead wish to use a dataset to optimize the prediction of one specific variable.
	The second is that the PCA process is very sensitive to the specific scaling of all the input variables.
	We will discuss this issue in more detail in Section~\ref{sec: data-pre-processing}

	For the reasons outlined above we chose to use a simple least-squares multiple linear regression method, as implemented in the Python package \texttt{Scikit-Learn} \citep{pedregosa2011}, to produce our relation.

\subsection{Data Preprocessing}\label{sec: data-pre-processing}

	To avoid biases and numerical issues with many multivariate analysis techniques it is important that the variables are of the same order of magnitude.
	The data therefore have to be standardized, which is often done by a process called whitening, i.e. subtracting the mean of each variable and dividing by the standard deviation.
	Given an intrinsically Gaussian distribution of the variable this operation projects each variable onto a Gaussian of mean equal to zero and standard deviation of one.
	While the Gaussianity assumption is not necessarily true, the conversion still ensures that the variable ranges are comparable.

	However, since this process is highly dependent on the specifics of the sample and not very generalizable outside it, we chose instead to make the value ranges comparable by moving into logarithmic space and subtracting a constant from them.
	This integer was chosen for each variable such that the remainder after subtraction is roughly order unity.
	This choice is somewhat arbitrary and therefore we examine any potential consequences of the standardization process in Section~\ref{sec: Standardization sensitivity} but find that it does not impact our results.

	We also note that since we choose to work in logarithmic space we cannot deal with negative values of variables. This limits us to galaxies with net Ly$\alpha$ emission.

\subsection{Error estimation}
	We use a Monte Carlo approach to estimate the uncertainty of the predictions that the multiple linear regression produces. We resample the data assuming that the measurement errors, given in Tables \ref{tab: Obs data}, \ref{tab: Obs data 2} and \ref{tab: Phys data}, are independent and Gaussian. We then redo the fit on the resampled data and record the predicted value for each galaxy. This process is repeated 1000 times and the uncertainty on the prediction for each galaxy is calculated as the range between the 16th and 84th percentiles of the 1000 values.

	There are several metrics available that can be used to quantify the performance of the model. The $R^2$ metric describes the fraction of the variance in a data set that is explained by a model. $R^2$ values thus range from 0 to 1 with 1 being a perfect explanatory model and 0 being no explanatory power at all. The mathematical definition of $R^2$ is as follows:
	\begin{equation}
	 R^2 = 1 - \frac{\sum_i(y_i-f_i)^2}{\sum_i(y_i - \bar{y})^2}
	\end{equation}
	where $y_i$ are the measured datapoints, $\bar{y}$ is the mean of the datapoints and $f_i$ is the model value corresponding to the datapoint. We also use root-mean-square error (RMS) as a metric of the dispersion around our relation since this is commonly used in the astronomical literature and  gives a quantitative measure of the overall uncertainty of the prediction.

\subsection{Cross validation}

	The goal of the current work is to find a multivariate relation that can be applied to other galaxy samples in order to predict their \lya{} luminosity. 
	We therefore need to quantify the performance of the relation when applied to galaxies that are not used in fitting the relation.
	We do this in two separate ways: first, we use a process known as cross validation which uses only the galaxies in the LARS sample; second, we attempt to gather a completely separate set of galaxies that we can apply the prediction to (see section \ref{sec: Out-of-sample}).

	Cross validation (CV) is a collective name for a group of techniques that all rely on leaving some objects out of a fit in order to use them as a separate testing group. 
	A common implementation is $k$-fold CV which splits the sample into $k$ groups, fits the model on $k - 1$ of the groups and tests the predictive ability on the remaining, excluded, group.
	This is then repeated until each group has been excluded from the fit once.
	The average predictive score then gives an estimate of the model's ability to generalize outside the fitted sample. We will use the RMS of the residuals as the metric for measuring CV results, however, since we also use this measure for the residuals of our full fit, we will refer to the results from CV as $CVS$ throughout to avoid confusion.

	When applying the $k$-fold CV methodology to a sample as small as ours however, some care must be taken.
	Since we have few galaxies in the sample the group-size must be kept small in order for there to be enough galaxies in the remaining sample to reliably fit the model.
	However, if the group size is too small the likelihood of outlier groups, i.e. an individual group that lies far from the mean distribution, becomes large, and such a group would have a large impact on the averaged CV score.
	Essentially the score becomes sensitive to the random selection of the $k$-fold grouping.
	We mitigate this effect by repeating the whole $k$-fold CV process, including group selection, 100 times and averaging the results of these runs using $k$=3.
	This samples the set of possible group selections and reduces the effect of any one instance that happens to have strong outlier groups.

\subsection{Variable selection}
	When we have a fitted relation, we also want to determine which variables hold the most predictive power.
	In principle, one could just look at the coefficients of the fitted function and say that the largest absolute values of the coefficients correspond to the most important variables.
	However, this assumes that the distributions and scaling of all variables are identical, which is not guaranteed.
	We therefore need another method for estimating the relative importance of the variables.
	The techniques we use are known as forward and backward selection. 

    The forward selection method works as follows.
    First we fit linear regressions between \lya{} luminosity and each individual variable in the data set.
    From these fits we calculate the $R^2$ and select whichever variable can explain the largest amount of variance as the most important.
    In the next step we add in each remaining variable to the data set in sequence and refit the regression, i.e. we fit \lya{} luminosity against 2 variables.
    The $R^2$ values are again computed and whichever variable increased the $R^2$ most is selected as the second most important variable.
    This is then repeated for 3 variables and so on.

    Backward selection works in the other direction.
    First the full relation is fit and the $R^2$ is calculated, then each variable is removed in turn, the relation is refitted and the $R^2$ recalculated.
    The variable whose removal reduced the $R^2$ the least is selected as the \emph{least} important variable.
    This variable is then removed from the dataset and the whole procedure is repeated until only one variable remains.

	We also perform a Monte Carlo simulation to determine the effect of observational noise on these rankings.  First we resample each datapoint using the observational error under the standard Gaussianity assumption. Then the variable selection, forwards and backwards, is done and the results are saved. This process is repeated 1000 times.


\section{Results}
      \subsection{Physical variables}\label{sec: phys results}
            After performing data standardization of the physical variable set, we fit a multidimensional linear relation to the whole dataset, using the \lya\ luminosity as the response variable 
            Since it is not possible to plot the full 9 dimensional relation, we instead show the predictions of our relation compared to the actual measured \lya\ in the top left panel of Figure \ref{fig: Fitted relations}.

            The model shown in the top left panel of Figure \ref{fig: Fitted relations} has an $R^2 = 0.85$ which means that it can explain 85\% of the variance of the observed \lya\ flux. As a reference point, the best individual variable---\lya\ correlation gives an $R^2$ of 0.48. This improvement provides strong support for the interpretation and treatment of \lya\ emission as a multivariate problem.

            We can also characterize the relation using the root mean square (RMS) of the residual as a measure of the dispersion of the points around the 1:1 relation, which in this case is 0.27 dex. The distribution of residuals is shown in the bottom left panel of Figure~\ref{fig: Fitted relations}, with the RMS range marked with the dashed lines.
            Using cross validation to quantify the predictive ability of this relation gives a $CVS$ of 0.39 dex.
            As we expect this is higher than the RMS of the fit but the difference is not large, which indicates that the relation generalizes outside the fitted sample quite well and that we are not significantly overfitting the galaxies included in the regression.

            The best fit coefficients and the corresponding variables are given in Table~\ref{tab: physvar formula}. The final relation is then given by 
            \begin{equation}
            \log(L_{\mathrm{Ly}\alpha}) -40  = \sum_{i=0}^n c_i \cdot v_i
            \label{eq: formula}
            \end{equation}
            where $c_i$ and $v_i$ is the $i$th coefficient and variable given in the table.

            \begin{deluxetable}{lll}[htb!]
                  \tablecaption{\label{tab: physvar formula}}
                  \tablecolumns{3}
                  \tablehead{
                  		\colhead{Index}&
                        \colhead{Coefficient} &
                        \colhead{Variable}
                  }
                  \startdata
                  0 & 0.443   	& log(M$_*$)$- 10$  \\
                  1 & -0.19	    	& log(UV size)     	\\
                  2 & -1.328        & E$_{B-V}$      	\\
                  3 & 0.159        & log(O32)          \\
                  4 & 0.130      	& 12+log(O/H)$- 7$  \\
                  5 & 0.843         & log(SFR)      	\\
                  6 & 0.673         & log(v$_{95}$)$-2$ \\
                  7 & -1.541 		& log(w$_{90}$)$-2$ \\
                  8 & -0.77		    &F$_{\mathrm{cov}}$				\\
                  \enddata
            \end{deluxetable}

            Table~\ref{tab: physvar variable selection} shows the results of forward and backward variable selection on this relation. We note that in this case the two methodologies agree very well which lends additional support for the assumption that this order reflects the actual information content of the variables. The distribution of the rankings after Monte Carlo is shown in Figure\,\ref{fig: Forward selection distribution}.

            \begin{deluxetable}{lll}[htb!]
                  \tablecaption{Resulting variable importance rankings for the forward and backward selection processes. The rankings range from 1 (Most important) to 9 (least important) 
                  \label{tab: physvar variable selection}}
                  \tablecolumns{3}
                  \tablehead{
                        \colhead{Ranking} &
                        \colhead{Forward selection} &
                        \colhead{Backward selection}
                  }
                  \startdata
                  1 & SFR      		& SFR      		\\
                  2 & E$_{B-V}$     & E$_{B-V}$     \\
                  3	& M$_*$			& M$_*$\\
                  4 & UV size       & UV size      \\
                  5 & F$_{\mathrm{cov}}$	     		& F$_{\mathrm{cov}}$	      \\
                  6 & O32      		& w$_{90}$      \\
                  7 & w$_{90}$      &v$_{95}$          \\
                  8 & v$_{95}$     	& O32     \\
                  9 & O/H       	& O/H      \\
                  \enddata
            \end{deluxetable}

    \subsection{Direct observables}
      	In the previous section we demonstrated that it is possible to predict the \lya\ luminosity from derived physical variables. However, working with direct observables has the benefit of removing any reliance on calibration relations that may or may not be applicable to the considered galaxies. We therefore use the same methodology to construct a relation for the observational data presented in Tables\,\ref{tab: Obs data} and \ref{tab: Obs data 2}.

\begin{figure*}[ht!]
                        \plotone{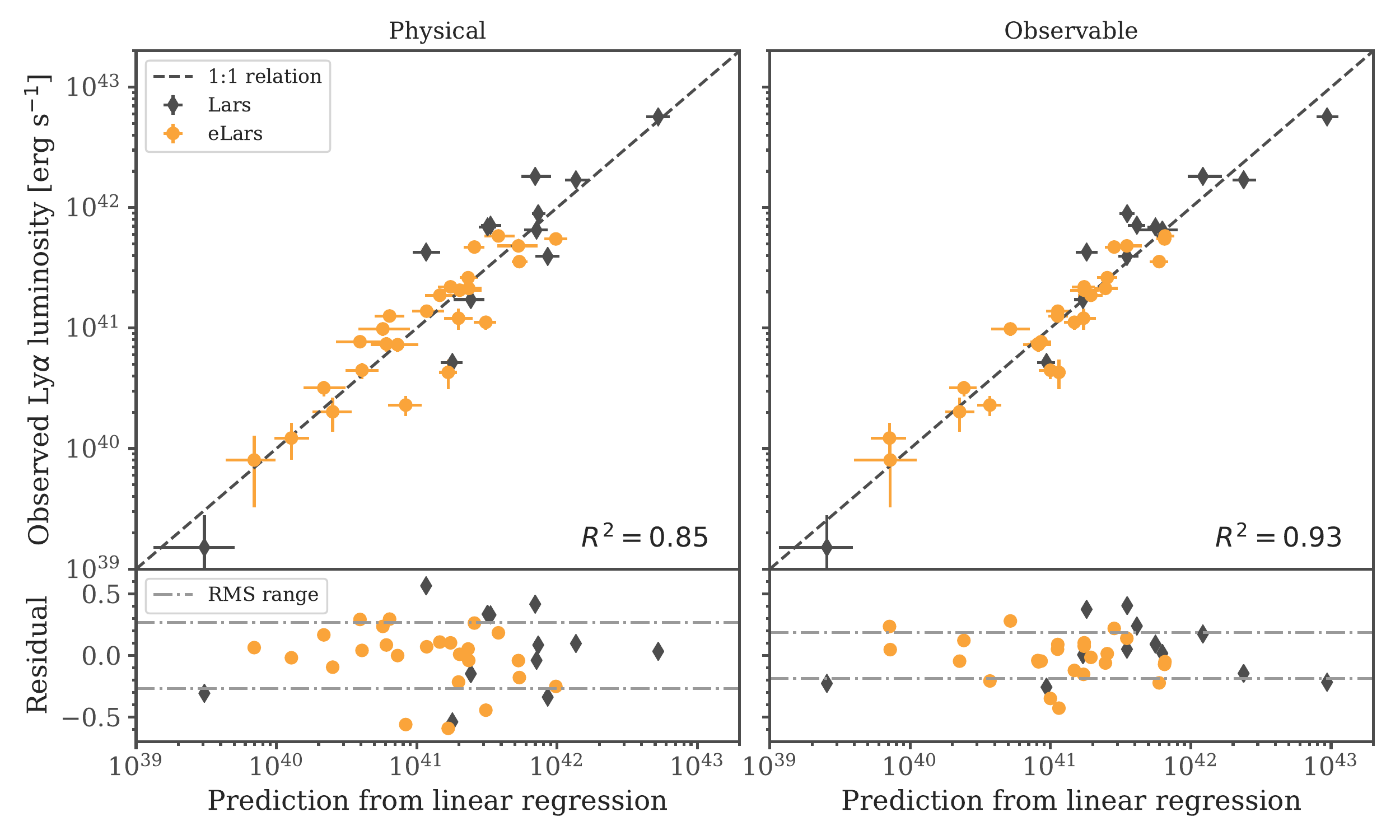}
                        \caption{The upper panels show the predicted \lya\ luminosity from our best fit relations versus the observed \lya\ luminosity for the physical (left) and observational (right) variable sets. The dashed line indicates the 1:1 relation expected from a perfect prediction. The lower panels show the residuals with the dash-dotted lines indicating $\pm$ the RMS of the relation. The black diamonds are galaxies from the original 14 LARS galaxies and the orange points are galaxies from the eLARS sample. \label{fig: Fitted relations}}
                  \end{figure*}

        The predictions from the fit to all the observational variables are shown in the right panels of Figure \ref{fig: Fitted relations}. The coefficients of the fit are shown in Table \ref{tab:_obsvar_formula} and can be used in Equation \ref{eq: formula} in the same way as the coefficients presented for the physical relation. 

            \begin{deluxetable}{lll}[htb!]
                  \tablecaption{Fitting coefficients and variables for the observable variable set. \label{tab:_obsvar_formula}}
                  \tablecolumns{3}
                  \tablehead{
                  		\colhead{Index}&
                        \colhead{Coefficient} &
                        \colhead{Variable}
                  }
                  \startdata
                  0 & -5.56    	& log(U)$-40$			\\
                  1 & 4.122	    	& log(B)$-40$     		\\
                  2 & -0.098       	& log(I)$-40$      		\\
                  3 & 2.538         & log(FUV)$-40$         \\
                  4 & -0.906      	& log(H$\alpha$)$-41$   \\
                  5 & -0.333        & log(UV size)      	\\
                  6 & 1.71         & log(H$\beta$)$-41$	\\
                  7 & 0.240 		& log(OIII)$-40$		\\
                  8 & -1.221	    & log(OII)$-40$			\\
                  9 & 0.503		& log(NII)$-40$			\\
                  10 & 0.063 		& v$_{95} / 100$		\\
                  11 & -0.093 		& w$_{90} / 100$		\\
                  12 & -0.359 		& F$_{\mathrm{cov}}$	\\
                  \enddata
            \end{deluxetable}

        Comparing the results in this figure to those in the left panel we note that there is some improvement. The new relation explains $\simeq 93\%$ of the variance in the data ($R^2=0.930$) and the residual has an RMS of 0.19 dex. This can also be seen by comparing the marked RMS ranges in the bottom panels of Figure~\ref{fig: Fitted relations}.

        We perform the cross validation of the results for the observational variables in exactly the same way as for the physical variable set and we find that the average $CVS$ after 100 sets of 3-fold cross-validation is $0.34 \pm 0.056$. 
        
\begin{figure*}[hbtp!]
            	\label{fig: Forward selection distribution}
                \plotone{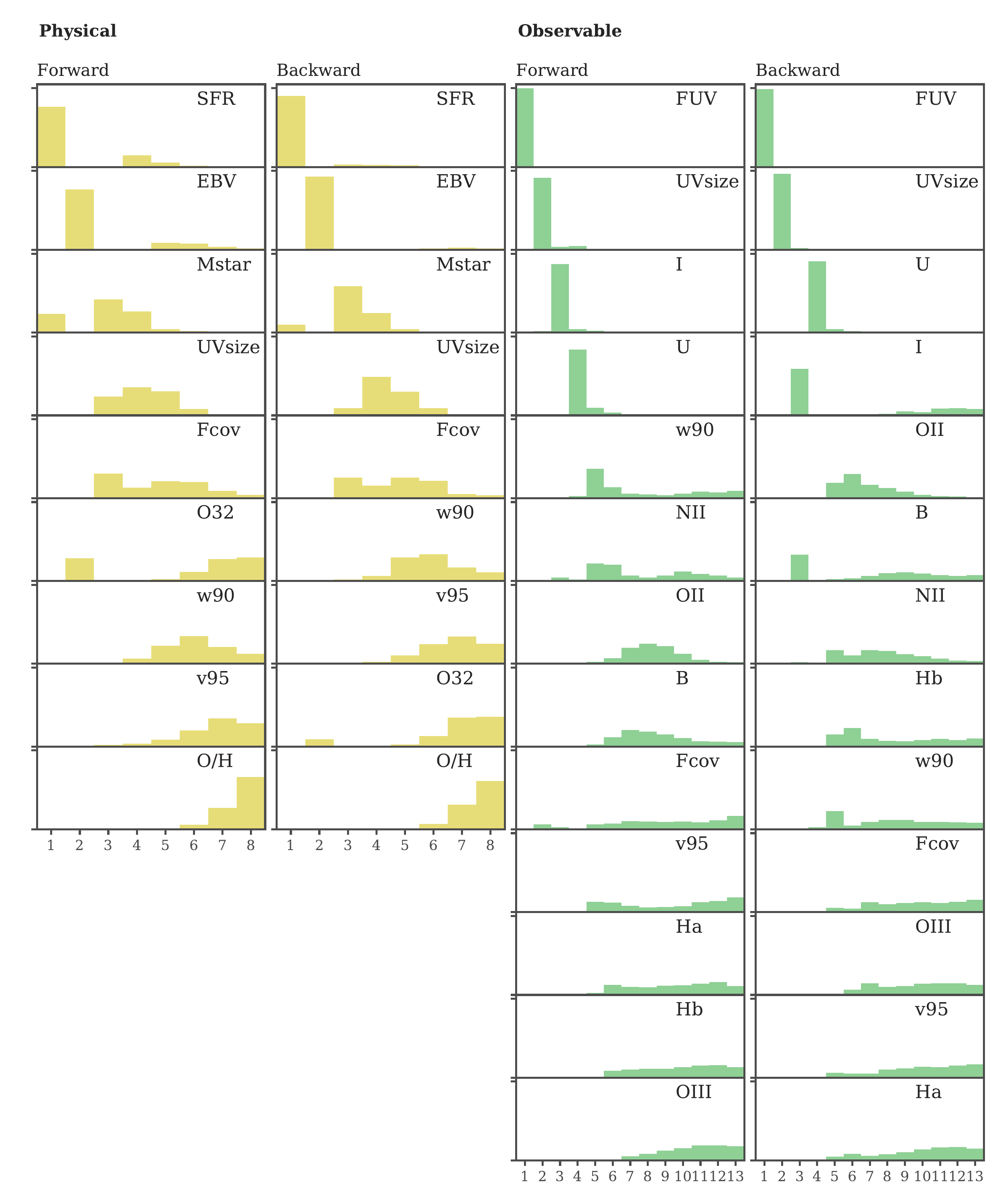}
                \caption{Distributions of rankings from a Monte Carlo simulation of the forward and backward selection process for both the physical (left) and observable (right) variable sets. The y--axis of all plots indicates relative frequency of occurrence. The variables are ordered by their average ranking, which is why the order does not exactly match Tables \ref{tab: physvar variable selection}, \ref{tab: obsvar variable selection}.\label{fig: selection distribution}}
            \end{figure*}

        This is consistent with the out-of-sample predictive ability of the relation we derived in Section~\ref{sec: phys results}. The fact that the difference between the RMS of  the fit and the $CVS$ is bigger for this relation than for the relation derived with physical variables is a slight indication of increased overfitting, which is expected based on the larger number of included variables. The increase is, however, still modest. We will look more into out-of-sample predictive ability in Section~\ref{sec: Out-of-sample} to make sure that we are not dominated by overfitting effects. The results of forward and backward selections are given in table \ref{tab: obsvar variable selection} and the distribution of rankings after Monte Carlo analysis is shown in Figure\, \ref{fig: Forward selection distribution}.
            \begin{deluxetable}{lll}[htb!]
                  \tablecaption{Resulting variable importance rankings for the forward and backward selection processes. The rankings range from 1 (Most important) to 13 (least important)\label{tab: obsvar variable selection}}
                  \tablecolumns{3}
                  \tablehead{
                        \colhead{Ranking} &
                        \colhead{Forward selection} &
                        \colhead{Backward selection}
                  }
                  \startdata
                  1 & FUV 			& FUV\\
                  2 & UV size		& UV size\\
                  3 & I				& U \\
                  4 & U 			& I \\
                  5 & w$_{90}$ 		& OII \\
                  6 & NII 			& B\\
                  7 & OII 			& NII\\
                  8 & B 			& H$\beta$ \\
                  9 & F$_{\mathrm{cov}}$	 			& w$_{90}$\\
                  10 & v$_{95}$ 	& F$_{\mathrm{cov}}$	\\
                  11 & H$\alpha$ 	& OIII\\
                  12 & H$\beta$		& v$_{95}$\\
                  13 & OIII 		& H$\alpha$\\
                  \enddata
            \end{deluxetable}

      \subsection{Assessing the stability of the results}
                   
            \subsubsection{Out-of-sample prediction}\label{sec: Out-of-sample}
                   The most intuitive way of testing whether our best-fit relation will yield reliable predictions when applied to other galaxy samples is to actually do this on a sample where the true \lya{} luminosity is known and our prediction can be tested. 
                   The sample of low to intermediate redshift ($0.1\lesssim z\lesssim 0.4$) starbursts assembled by \citet{yang2017} provides an appropriate and comparable dataset. The galaxies are compact starbursts with extensive COS observations by \citet{heckman2011,jaskot2014,henry2015,izotov2016}. The star-formation rates of these galaxies range between roughly 1 and 30 $M_\odot \mathrm{yr}^{-1}$, the masses are between $10^8$ and $10^9$ $M_\odot$ and the metallicities are $7.7\lesssim \log(O/H)+12 \lesssim 8.3$. These properties makes these galaxies comparable to a subset of the LARS galaxies and hence ideal for use for testing our relations. 

                   We present the individual galaxies, along with the original GO proposals and the compiled data in Table \ref{tab: GP data}. We will refer to this dataset as the test sample henceforth.

                \begin{longrotatetable}
                \begin{deluxetable*}{lllllllllllllllll}
					\movetabledown=2in
                  \tablecaption{Data collected for the compact starburst test sample.\label{tab: GP data}}
                  \tablecolumns{16}
				\tablehead{
					\colhead{ID}& 
					\colhead{GO}& 
					\colhead{Ly$\alpha$ \tablenotemark{1}}& 
					\colhead{U }& 
					\colhead{B }& 
					\colhead{I }& 
					\colhead{FUV  \tablenotemark{1}}& 
					\colhead{H$\alpha$ }& 
					\colhead{UVsize \tablenotemark{1}}& 
					\colhead{H$\beta$ }& 
					\colhead{[OIII] }& 
					\colhead{[OII] }& 
					\colhead{[NII] }& 
					\colhead{v95 }& 
					\colhead{w90 }& 
					\colhead{F$_\mathrm{cov}$} \\
					\colhead{}& 
					\colhead{}& 
					\colhead{$10^{41}$}& 
					\colhead{$10^{40}$}& 
					\colhead{$10^{40}$}& 
					\colhead{$10^{40}$}& 
					\colhead{$10^{40}$ \tablenotemark{1}}& 
					\colhead{$10^{40}$}& 
					\colhead{}& 
					\colhead{$10^{40}$}& 
					\colhead{$10^{40}$}& 
					\colhead{$10^{40}$}& 
					\colhead{$10^{40}$}& 
					\colhead{}& 
					\colhead{}& 
					\colhead{}& 
				}
				\startdata
				0021+0052 &	13017 &	$39.81$ &	$1.71$ &	$1.64$ &	$0.49$ &	$12.14$ &	$281.81$ &	$0.44$ &	$97.84$ &	$462.32$ &	$92.32$ &	$21.62$ &	$924.01$ &	$859.26$ &	$0.74$\\ 
				0749+3337 &	14201 &	$2.0$ &	$4.05$ &	$4.19$ &	$1.82$ &	$2.24$ &	$1.06\cdot10^{3}$ &	$1.47$ &	$366.08$ &	$1.38\cdot10^{3}$ &	$498.84$ &	$116.7$ &	$399.81$ &	$566.32$ &	$0.03$\\ 
				0815+2156 &	13293 &	$19.95$ &	$0.39$ &	$0.29$ &	$0.12$ &	$2.43$ &	$72.4$ &	$0.35$ &	$25.27$ &	$186.54$ &	$7.69$ &	$1.11$ &	$513.03$ &	$995.3$ &	$0.77$\\ 
				0822+2241 &	14201 &	$19.95$ &	$1.46$ &	$1.47$ &	$0.74$ &	$3.87$ &	$781.8$ &	$0.68$ &	$268.25$ &	$1.56\cdot10^{3}$ &	$267.76$ &	$48.02$ &	$473.98$ &	$501.97$ &	$0.17$\\ 
				0911+1831 &	12928 &	$63.1$ &	$2.6$ &	$1.93$ &	$0.88$ &	$11.17$ &	$435.49$ &	$0.57$ &	$151.07$ &	$544.72$ &	$159.5$ &	$52.6$ &	$726.35$ &	$659.85$ &	$0.45$\\ 
				0917+3152 &	14201 &	$50.12$ &	$3.73$ &	$2.59$ &	$0.92$ &	$13.19$ &	$322.43$ &	$0.47$ &	$112.15$ &	$393.54$ &	$101.31$ &	$46.41$ &	$566.4$ &	$625.58$ &	$0.48$\\ 
				0938+5428 &	11727 &	$3.16$ &	$1.35$ &	$1.14$ &	$0.45$ &	$9.04$ &	$239.03$ &	$0.47$ &	$82.64$ &	$361.95$ &	$78.9$ &	$19.6$ &	$503.7$ &	$654.56$ &	$0.25$\\ 
				1018+4106 &	14201 &	$7.94$ &	$0.89$ &	$0.65$ &	$0.3$ &	$2.4$ &	$208.75$ &	$0.78$ &	$72.33$ &	$450.81$ &	$49.84$ &	$7.94$ &	$532.82$ &	$612.93$ &	$0.18$\\ 
				1025+3622 &	13017 &	$19.95$ &	$1.29$ &	$1.13$ &	$0.4$ &	$7.59$ &	$157.77$ &	$0.76$ &	$54.86$ &	$284.79$ &	$46.53$ &	$8.49$ &	$394.76$ &	$394.25$ &	$0.4$\\ 
				1032+2717 &	14201 &	$2.0$ &	$0.98$ &	$0.87$ &	$0.28$ &	$3.63$ &	$6.76$ &	$0.63$ &	$2.47$ &	$8.91$ &	$0.44$ &	$1.37$ &	$273.38$ &	$372.28$ &	$0.24$\\ 
				1054+5238 &	12928 &	$31.62$ &	$5.19$ &	$3.67$ &	$1.33$ &	$17.87$ &	$534.31$ &	$0.62$ &	$185.83$ &	$862.25$ &	$188.7$ &	$37.5$ &	$628.59$ &	$635.14$ &	$0.51$\\ 
				1122+6154 &	14201 &	$15.85$ &	$0.53$ &	$0.39$ &	$0.15$ &	$2.64$ &	$115.01$ &	$0.32$ &	$39.89$ &	$224.65$ &	$27.79$ &	$6.05$ &	$274.92$ &	$383.93$ &	$0.4$\\ 
				1133+6514 &	12928 &	$39.81$ &	$1.46$ &	$0.97$ &	$0.41$ &	$9.41$ &	$128.45$ &	$0.82$ &	$44.81$ &	$256.61$ &	$29.88$ &	$4.9$ &	$438.62$ &	$387.51$ &	$0.78$\\ 
				1137+3524 &	12928 &	$39.81$ &	$2.36$ &	$2.1$ &	$0.74$ &	$9.85$ &	$383.13$ &	$0.72$ &	$133.15$ &	$686.81$ &	$129.81$ &	$25.72$ &	$439.94$ &	$464.87$ &	$0.31$\\ 
				1219+1526 &	12928 &	$158.49$ &	$1.37$ &	$1.12$ &	$0.38$ &	$9.63$ &	$276.91$ &	$0.33$ &	$96.58$ &	$614.62$ &	$30.79$ &	$8.51$ &	$834.89$ &	$820.44$ &	$0.75$\\ 
				1244+0216 &	12928 &	$31.62$ &	$2.44$ &	$2.41$ &	$0.93$ &	$6.73$ &	$699.49$ &	$1.02$ &	$242.6$ &	$1.45\cdot10^{3}$ &	$214.85$ &	$30.63$ &	$341.25$ &	$403.07$ &	$0.19$\\ 
				1249+1234 &	12928 &	$125.89$ &	$2.01$ &	$1.45$ &	$0.51$ &	$12.37$ &	$364.35$ &	$0.71$ &	$126.51$ &	$778.28$ &	$112.47$ &	$12.15$ &	$556.39$ &	$542.02$ &	$0.49$\\ 
				1339+1516 &	14201 &	$7.94$ &	$0.99$ &	$0.87$ &	$0.41$ &	$1.78$ &	$334.59$ &	$0.38$ &	$115.96$ &	$745.22$ &	$61.01$ &	$15.48$ &	$376.98$ &	$450.01$ &	$0.2$\\ 
				1424+4217 &	12928 &	$79.43$ &	$1.89$ &	$1.54$ &	$0.49$ &	$8.88$ &	$394.21$ &	$0.48$ &	$137.13$ &	$894.17$ &	$78.92$ &	$16.6$ &	$532.35$ &	$533.93$ &	$0.55$\\ 
				1428+1653 &	13017&	$31.62$ &	$2.35$ &	$2.55$ &	$1.06$ &	$10.87$ &	$357.59$ &	$0.77$ &	$123.14$ &	$372.5$ &	$155.26$ &	$49.46$ &	$399.63$ &	$391.5$ &	$0.46$\\ 
				1429+0643 &	13017 &	$50.12$ &	$2.82$ &	$2.43$ &	$0.86$ &	$11.74$ &	$604.91$ &	$0.4$ &	$209.99$ &	$1.27\cdot10^{3}$ &	$148.86$ &	$40.52$ &	$721.34$ &	$756.82$ &	$0.39$\\ 
				1440+4619 &	14201 &	$63.1$ &	$6.49$ &	$4.38$ &	$1.84$ &	$18.67$ &	$624.64$ &	$0.72$ &	$216.89$ &	$953.45$ &	$256.87$ &	$45.89$ &	$895.06$ &	$824.43$ &	$0.27$\\ 
				1457+2232 &	13293 &	$2.0$ &	$0.71$ &	$0.64$ &	$0.26$ &	$3.76$ &	$194.76$ &	$0.42$ &	$67.7$ &	$504.19$ &	$29.88$ &	$3.83$ &	$443.39$ &	$486.65$ &	$0.21$\\ 
				1543+3446 &	14201 &	$1.0$ &	$0.37$ &	$0.32$ &	$0.13$ &	$1.85$ &	$40.24$ &	$0.77$ &	$14.15$ &	$76.97$ &	$5.71$ &	$1.2$ &	$268.37$ &	$607.64$ &	$0.21$\\ 
				2237+1336 &	14201 &	$12.59$ &	$3.45$ &	$2.8$ &	$1.2$ &	$8.23$ &	$512.2$ &	$1.08$ &	$177.69$ &	$853.15$ &	$218.52$ &	$25.2$ &	$491.01$ &	$528.82$ &	$0.26$\\ 
				\enddata
                  \tablenotetext{1}{From \citet{yang2017}}
            	\end{deluxetable*}
            	\end{longrotatetable}
                We chose these galaxies as a comparison sample for a number of reasons. Firstly, they have all the observations required for us to be able to test our predictions on them. Secondly, they mostly lie inside the parameter space covered by our galaxies for instance in terms of a number of properties such as \lya\ and \halpha\ luminosity. In fact LARS 14  is also included in their sample (J0926+4427) but we retain it in our prediction set. This rough overlap means that we can expect our relation to be able to give reasonable predictions for the test sample.

                The specific galaxies used here are selected by the classic Green Pea selection of strong [\oIII]\,4959\,, 5007 and \hbeta\ emission (\citet{cardamone2009} but see \citet{yang2017} for details on these specific galaxies). This selection means that the galaxies lie in the upper end of the star formation rates covered by the LARS, but also that the galaxies are more uniform in properties than our original sample since extreme [\oIII{}] equivalent widths ensures selection of galaxies with relatively low metallicities, and low E$_{B-V}$ and high ionization parameter (O$_{32}$ ratio).

                The full dataset for this sample was collected from several different sources.
                We retrieved the UV sizes along with the \lya{} luminosity from the tabulated data in \citet{yang2017} and calculated the FUV luminosity from the tabulated \lya{} flux and equivalent widths in that paper.
                We remeasured the optical line luminosities identically to the procedure we used for LARS; simultaneously fitting Gaussians to a running median continuum subtracted SDSS spectrum.

                For the test data, we also require U, B, and I magnitudes that we use for the prediction based on observable quantities.
                Therefore, we used the package \texttt{pysynphot} to create synthetic photometric measurements for the B and I bands from the SDSS spectrum.
                This synthetic photometry should be comparable to the large aperture photometry of the LARS galaxies due to the  compactness of the galaxies in this sample.
                The U band (F336W) however lies outside the spectral coverage of the SDSS and we therefore had to approximate the HST band with the SDSS u band flux.

                  \begin{figure*}[ht!]
                        \plotone{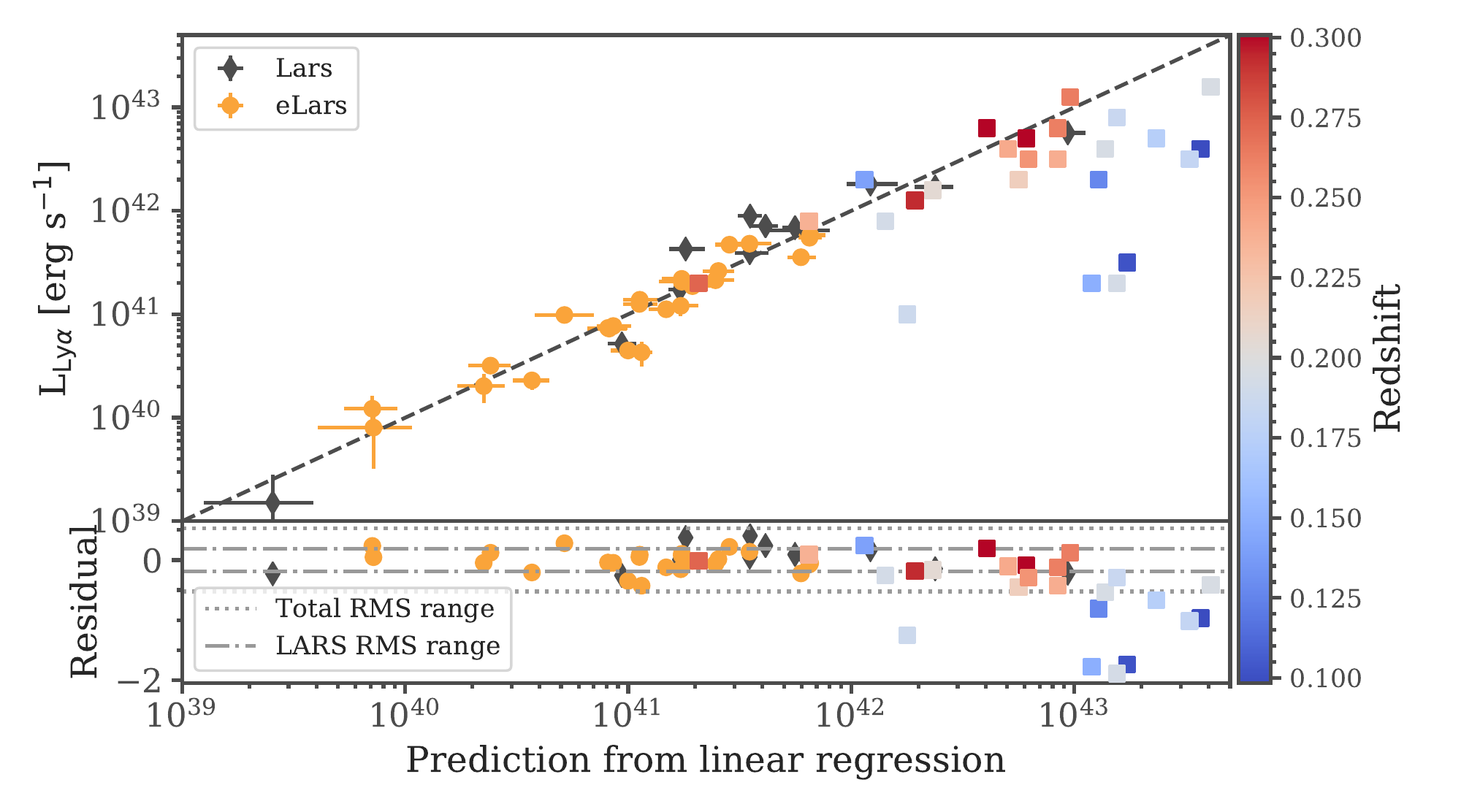}
                        \caption{The predicted \lya{}  from our best fit relation for both the LARS and test samples versus the observed \lya{} luminosity. The dashed line indicates the 1:1 relation expected from a perfect prediction. The coloration of the test sample indicates the redshift of the galaxy.
                        The bottom panel shows the residuals as well as the RMS range calculated from the LARS galaxies (dash-dotted line) and all of the galaxies (dotted line)\label{fig: GP relation}}
                  \end{figure*}

                  Now that we have acquired a complete comparable dataset for the test sample we use our previously derived relation to predict the \lya{} luminosity of these galaxies.
                  This is shown in Figure \ref{fig: GP relation}.

                  We see that our relation in general performs very well at predicting the \lya{} luminosities of the test sample.
                  We note that there are four galaxies that out-lie the relation by a significant amount while the rest of the test sample follows our prediction reasonably well, but does show a slight deviation towards over-prediction, especially at the higher luminosities. It is likely that this is an effect of our fit being less constrained in that parameter region due to the dearth of such luminous galaxies in the LARS sample. It is also possible that aperture effects are important which we discuss in more detail in Section \ref{sec: outliers}.

            \subsubsection{Sensitivity to standardization of data}\label{sec: Standardization sensitivity}
                 \begin{figure}[hbt!]
                        \plotone{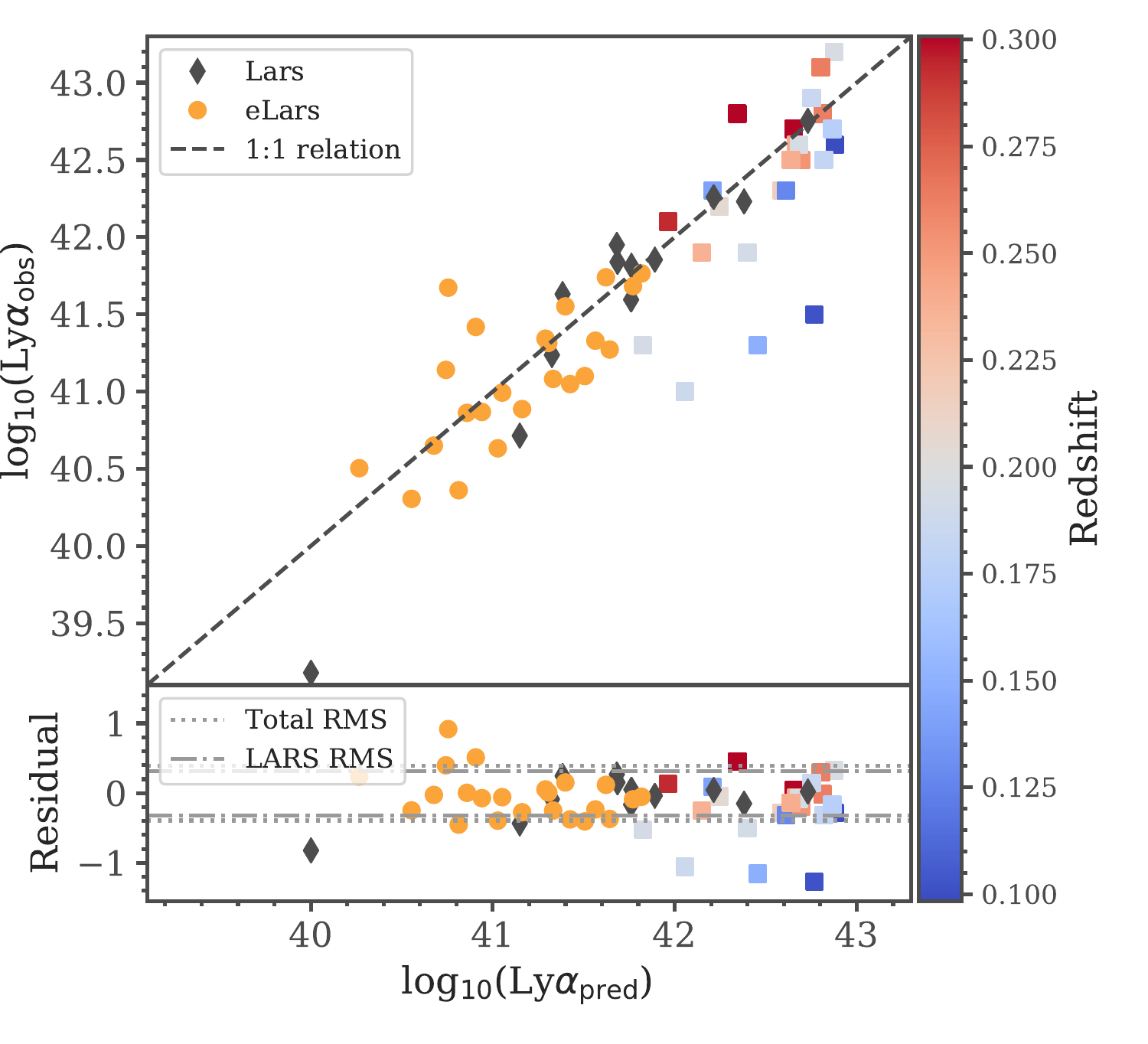}
                        \caption{Prediction of the emitted \lya{}  on the x-axis versus the observed \lya{} luminosity on the y-axis. The fitting was performed in linear space on the observable variable set and the result was subsequently logged, to more clearly show the distribution of the data. 
                        The coloration of the test sample indicates the redshift of the galaxy.
                        \label{fig: Linear fit} }
                  \end{figure}
                  When we standardize our variables we subtract a constant from each logged variable so as to make the final value approximately centered on one. As we stated before, the arbitrary nature of the normalization constants  makes it crucial to examine their impact on our results in more detail. We therefore performed some tests which consisted of altering the constants in various, more or less drastic ways and then examining the effect on our predictions. The first tests simply consisted of increasing and decreasing the constants by 2, which had no effect on our predictions. We then proceeded to more drastic tests such as adding $10^{10}$ to all variables in linear space, and removing the constant altogether. Neither of these had any effect. From this we conclude that the exact method and fitting routine that we use is very insensitive to these numerical effects.
                  
                \subsubsection{Logarithmic or Linear space}
                  The robustness of the fitting method leads to the follow-up question of whether it would be possible to do the fitting in linear space. This would have some advantages, such as enabling us to incorporate galaxies that have net absorption of \lya{} radiation in the sample. We did this by directly performing the multiple linear regression on the unstandardized observable dataset. The result, shown in Figure \ref{fig: Linear fit}, is encouraging since it clearly shows that the prediction is on average quite good.
                                    
                  When we quantify the dispersion however it is clear that the RMS of the linear fit (0.32 dex) is higher than the log fit (0.19 dex) and that there are more outliers to the relation in than in the logarithmic case. We conclude that it is possible to extract most of the predictive information in a linear regime but there is a risk of the information being drowned by numerical effects coming from the variable scales being widely disparate. Another important limitation that enters is the functional form: a linear fit is a more constrained model than a powerlaw which is what a linear regression in log space produces. 

                  The Figure also shows the predictions of the test sample from the linear space fit. These points actually show a slight improvement compared to the log fit, in that they seem to cluster more around the line. There are however still some distinct outliers. Our conclusion from this is that even though a linear fit is possible, a fit in logarithmic space is preferred and that the most likely cause for this is the somewhat less restrictive functional form that this fit implies.

\section{Discussion}
	\subsection{The physical variable relation}
		The relation we derived in Section \ref{sec: phys results} shows that it is possible to predict \lya{} emission and also very clearly shows that it can be well treated by even a relatively simple multivariate regression. We also show that we can reliably determine the order of importance of the constituent variables for this relation. 

		The most important variables that we determine agree reasonably well with expectations from theory of what should have important impacts. As we discussed in the introduction star formation plays a crucial role in the production of ionizing photons and therefore in production of \lya\ \citep{dijkstra2014}. We independently recover this dependence and see that star formation rate proved the most crucial predictor of \lya\ emission and carries a strong positive weight in the fit. Previous results, such as \citet{verhamme2006a,scarlata2009,schaerer2011,dijkstra2014}, have also indicated the importance of dust and how it should supress the emission of \lya\ which is indeed what we find. E$_{B-V}$ is the second most important variable and has a negative weight in the fit, indicating an anticorrelation as expected as more dust is available to absorb the \lya{} radiation. 
		The third most important variable is the stellar mass, where the connection to \lya\ emission is less obvious. The most probable cause is that a more luminous galaxy tends to be more luminous in all wavelengths, and the mass-luminosity relation relates the mass to this overall normalization.

	\subsection{Observable variable relation}
		The relation derived from direct observables clearly shows an improvement compared to that derived from physical quantities and to our knowledge also shows the best ability to predict \lya{} of any empirical relation in the literature. Testing the predictive potential of the relation on a completely separate sample also shows remarkable performance. The measurements for the test sample and the LARS sample differ in some key respects. Firstly in the measurements of the \lya{} and FUV luminosities. In the LARS sample the \lya{} luminosity is derived from a very large aperture measurement in a narrowband image, whereas for the test sample galaxies it is estimated from a numerical integration of the spectrum measured in the relatively small COS aperture. Similar considerations apply for the FUV luminosity which for LARS is a broadband measurement and for the testing sample sample it is estimated as the continuum level around \lya{} in the COS spectrum. There are differences of the same nature in the other broadband luminosities that we use. The LARS data uses HST imaging which we approximated for the test sample using synthetic photometry based on the SDSS spectrum, or, in the case of the U-band, the nearest SDSS broadband filter. However, the compactness of the Green Pea galaxies should mean that most of the flux emitted will be captured in the SDSS fiber, with the possible exception of  \lya{} which is expected to be more spatially extended.

		All of these measurement differences could be expected to introduce scatter, or systematic differences when we attempt to predict the test sample \lya{} from a model fitted to the LARS sample. Nevertheless we observe that the model performs relatively well, and produces an RMS$\sim0.5$ for the test sample predictions which is slightly larger than what we expected from cross validation but not inconsistent with this result. This robustness to technical differences is a key property that greatly strengthens our confidence in the usability of the relation on other galaxy samples, including those at high redshift.
		
		\subsection{Explaining the outliers}\label{sec: outliers}
		The observable relation we derive holds more predictive power than the physical relation and the majority of galaxies in this sample lie close to the 1:1 line, but there are four clear outliers, 1513+3446, 1032+2717, 1457+2232, and 0938+5428, that are underluminous in their measured \lya\ compared to the prediction.  Indeed, all four fall $\gtrsim$1 dex below the line, and are more discrepant than any of the (e)LARS galaxies used for the prediction. 

		We first note that these galaxies are 4 of the 5 lowest equivalent width galaxies in the sample, and also all show damped absorption wings around the \lya{} line. This makes the continuum level difficult to define and therefore adds significant uncertainty to the determination of their total \lya{} luminosity. 
		Another potential reason could be that the COS observations do not capture all the \lya{} emission from these galaxies. 

		\citet{henry2015} compared the COS spectroscopy to the \lya{} luminosity of LARS 14 published in \citet{hayes2014} and found a factor of $2$ difference. However, the zero point calibration of the SBC detector has since been updated \citep{avila2019} and the new total \lya{} luminosity from imaging is only 5\% larger than that measured in COS, despite the differences in aperture.

		In general, however, the spatially extended nature of \lya{} emission makes global luminosity measurements sensitive to aperture effects. Intuitively one would expect the test sample galaxies to have larger aperture losses due to the small COS aperture and the flux losses due to vignetting. At redshift 0.1, the lowest redshift of the testing sample, the radius of the COS aperture (1.25'') corresponds to a physical size of 2.3 kpc. The median scale length of the halos in the LARS sample is 3.38 kpc (Rasekh et al. in prep) and we could therefore expect some significant losses for the low redshift test galaxies. In contrast, at redshift 0.3 (the high-$z$ end of the testing sample) the  COS aperture covers a physical radius of 5.56 kpc which means that aperture losses should be almost negligible, assuming the scale length of these halos are comparable to LARS.

		In figure \ref{fig: GP relation} the test galaxies are color--coded by their redshift, and there appears to be a trend, with galaxies with redshifts lower than 0+.2 showing a systematic underestimation of their \lya\ compared objects with $z>0.2$. This provides strong support for the conclusion that aperture effect are of greater importance for the low-z test sample galaxies and that this is, at least partly, the reason for the fact that they diverge from our predictive relation.
		Detailed studies of the sizes of the LARS and eLARS halos show that there is large variation in the sample of the \lya{} halo spatial extent (Rasekh et al, in prep) which means making a correction is not possible since the aperture effect size will most likely vary strongly from object to object.

		\subsection{Predicting other Lyman alpha properties}
			From the variable selection process it is clear that a large fraction of the predictive power in our relation comes from one dominant variable. In the case of the observables it is the FUV luminosity and in the case of the physical variables it is the SFR. This is not surprising since the FUV luminosity is one of the most commonly used SFR indicators. One can then attempt to fold these first order relations out and determine to what extent the other variables impact the prediction. 

			The \lya\ and FUV luminosity combine easily into one of the most common observed quantities of \lya\ emission which is the \lya\ equivalent width. If the FUV luminosity were the only factor that determined the Ly$\alpha$ luminosity we would observe constant Ly$\alpha$ equivalent width for all galaxies. This is clearly not the case, and it is therefore interesting to see how well we can capture the deviations from this approximation. 

			The same reasoning can be applied to the physical variables where the SFR dominates. The SFR is calculated from the dust corrected \halpha\ luminosity. Since \halpha\ is also produced by hydrogen recombinations but escapes freely after emission, that luminosity can be combined with the \lya\ luminosity to measure the escape fraction of \lya{}. Again we can fold out the dominant predictor and instead create a relation for the \lya\ escape fraction. 

			\subsubsection{Equivalent widths}
\begin{figure}[ht!]
                    \plotone{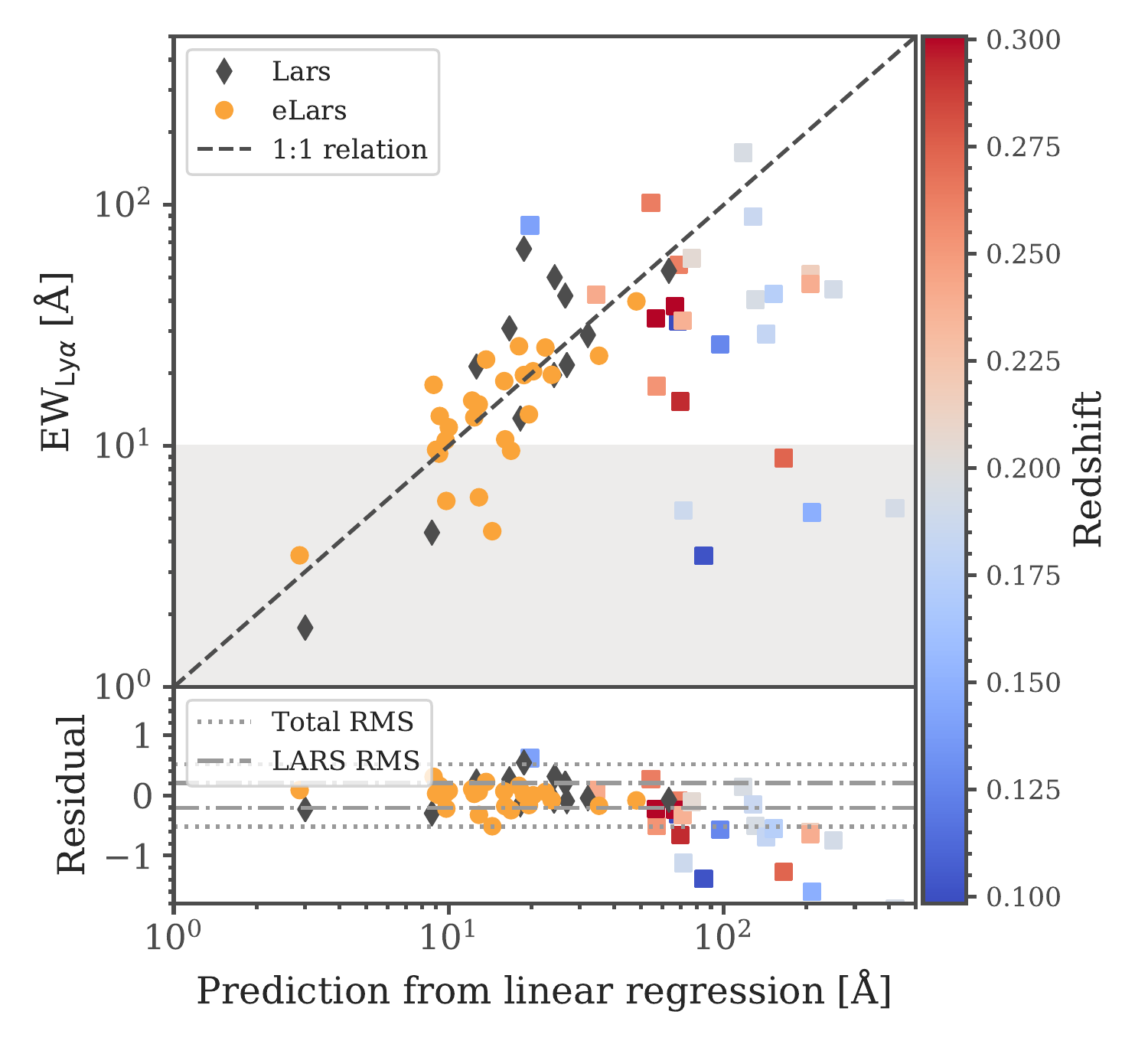}
                    \caption{The predicted EW for the test sample and the LARS sample from a fit to only the LARS sample. The shaded region indicates an observed equivalent width less than 10. \label{fig: EW relation}}
            \end{figure}

			The results of predicting the equivalent widths are shown in Figure \ref{fig: EW relation}. The only difference from our main relation is that the response variable is the equivalent width and that the FUV luminosity is not included as a predictor; all other things are kept the same.

            Figure \ref{fig: EW relation} shows that it is indeed possible to predict the EWs quite well although the new prediction has significantly larger scatter around the 1:1 relation than the luminosity prediction. This is expected since we saw that the dominant predictive power was given by the FUV luminosity. It is worth noting that the prediction appears to be worst for the lowest observed equivalent widths. Comparing Figure \ref{fig: EW relation} to  Figure \ref{fig: GP relation} and \ref{fig: Linear fit} show that they all display similar trends with redshift for the testing sample, although it is less clear for the EW prediction which is consistent with the larger scatter . It is also clear that the for the testing sample, the relation holds less predictive power at these very low EW$\leq10$ \AA\ (shaded region in the figure). 

            However, when \lya\ emitters (LAEs) are selected in narrowband or IFU surveys there is a threshold in EW (commonly 20Å) below which the galaxies are not considered LAEs. This is due to observational reasons, specifically the difficulty of obtaining sufficiently deep continuum observations for fainter emitters. Holding predictive power at very low EWs may therefore not be important for the application of the relation to higher redshift samples. We also note that the four outlier galaxies that we discussed in section \ref{sec: outliers} are all below the EW cut. For these reasons we remove galaxies with EW$\leq10$ \AA\  from the analysis and focus on predicting the higher EWs. For this subsample the RMS of the relation for all the galaxies is 0.24 dex. This is still substantial given the fact that the dynamic range is roughly one dex, but there is some predictive power in this relation. The $R^2$ value for the relation is only 0.18 for the total sample of galaxies but 0.55 for the LARS meaning 55\,\% of the total variance is explained by the fit. This indicates that this relation does not generalize as well as the main relation.

            We also attempt to do variable selection on this relation to see if we can establish a clear variable importance but we find that the backward and forward selection diverge from each other. This means it is not possible to constrain the most powerful predictors and that we need additional data in order to make a strong statement on this issue. 

            \subsubsection{Escape fraction}
\begin{figure}[h!]
                    \plotone{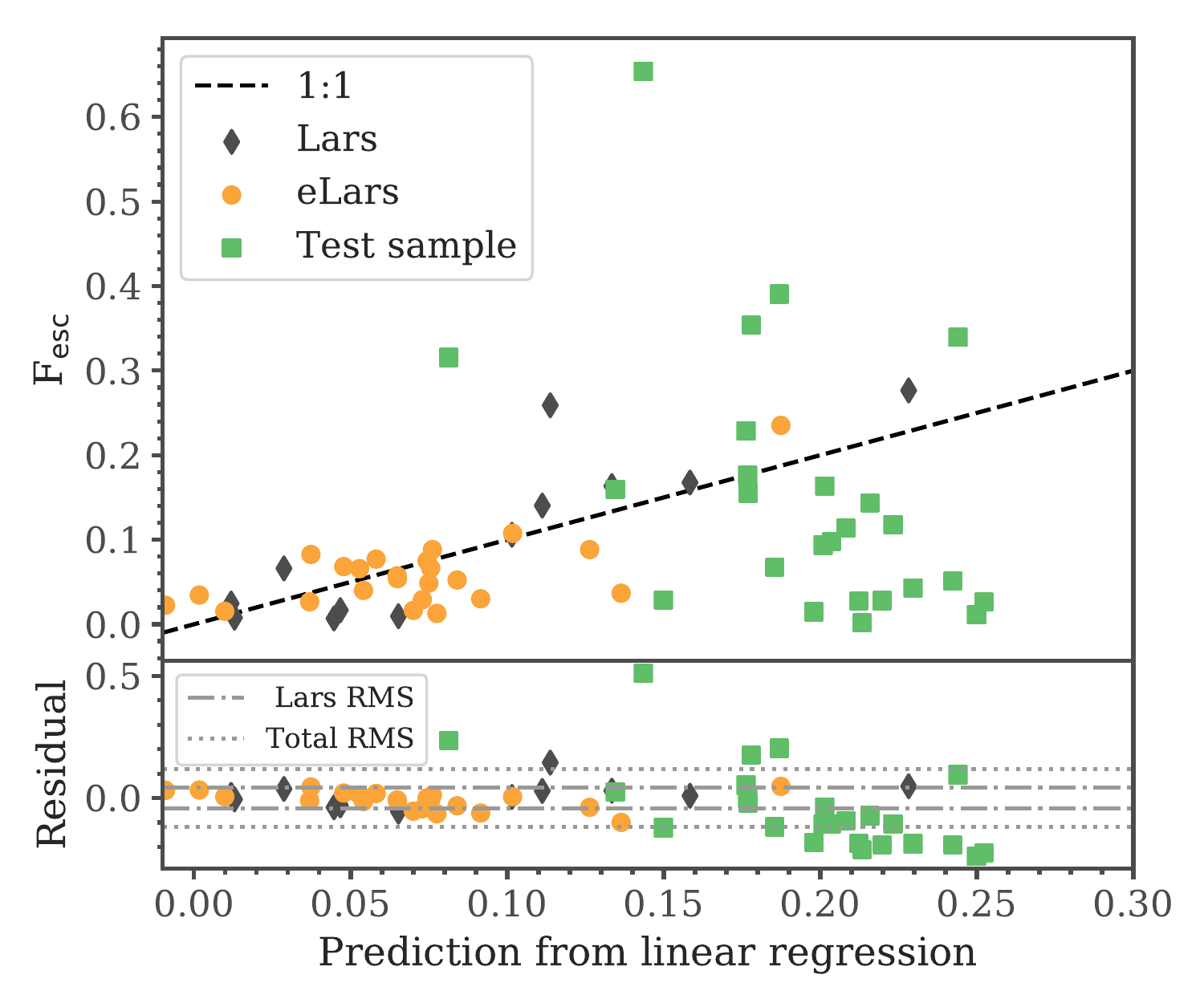}
                    \caption{The predicted $F_\mathrm{esc}$ for the LARS sample.
                    \label{fig: Fesc relation}}
            \end{figure}

            The escape fraction of \lya{} is derived by comparing the observed \lya{} and the intrinsic H$\alpha \times 8.7$ where the factor 8.7 comes from the intrinsic line ratios in a case B recombination scenario. This factor is an uncertain assumption however (see for instance footnote 11 in \citet{henry2015}). The results of predicting escape fractions are shown in Figure~\ref{fig: Fesc relation}.
       		The $R^2$ of this fit is 0.60 and the RMS$=0.04$ for the LARS and eLARS samples. However, the RMS for the testing sample is 0.8 which indicates that the relation does not generalize very well. There may be several reasons for this. The first is that the physical variable data for the testing sample are not calculated with the same methods. In particular the stellar mass determinations for the testing sample that enter this prediction are from \citet{yang2017} who used the measurements of \citet{izotov2011} and the MPA-JHU SDSS catalog. Both of sets are measured using SED modelling of the SDSS optical spectrum, i.e. a global SED modeling procedure, quite unlike the pixel-wise SED modeling used for the LARS galaxies. Another reason for the large scatter may be the lack of dynamical range in escape fraction in the lars sample. As we can see the LARS galaxies all have escape fractions below 30\% whereas the test sample have escape fractions as high as 66\% (1219+1526).
            \subsubsection{Spectral Ly$\alpha$ properties}\label{sec: spectral properties}
    		Given the success of our modelling at predicting global \lya{} properties we can also attempt to predict more specific quantities, such as the velocity shift of the red peak. This size of velocity shift is related to the \lya{} peak separation which has been claimed to be strongly correlated with the escape of Lyman continuum radiation \citep{izotov2018} and also impacts the optical depth that the \lya{} experiences as it travels through a neutral IGM. It would therefore be interesting if we could predict this quantity as well. For this reason we went through our sample of galaxies and selected the ones where a clear red peak was visible, leaving us with 24 galaxies for which we measured the peak position. We then fitted our relation to these data. What we find when doing comparisons with external samples is that, even though the initial results appear promising ($R^2=0.68$ within the LARS sample), we are dominated by overfitting effects due to the low number of galaxies.

    	\subsection{Comparison to Trainor et al. 2019}\label{sec: Trainor}

			A recent publication by \citet{trainor2019} studied a much larger (377 in the most relevant example) sample of $\langle$z$\rangle$=2.3 galaxies and reported
			that they could capture $\approx 90$~\% of the variance in the \lya\ EW
			using a smaller number of parameters.  Their preferred predictive variable was a non-linear combination of the EW of low ionization stage (LIS)
			absorption lines in the UV and the O3 ratio ($\equiv$~log([\oIII]5007/\hbeta)):
			$$X_{\mathrm{LIS}}^{O3} = \alpha(\mathrm{EW_{\mathrm{LIS}}}) + (1-\alpha)\mathrm{O3}$$
			Their result seems comparable with the primary results presented here (93\% of variance explained). We note that Trainor et al. predict \lya\ EW and, compared to our EW prediction, their relation does appear more encapsulating. However, there are some general differences in the employed methodologies that must be borne in mind when comparing their results to ours.

			Firstly, their definition of variance capture includes observational
			errors in two dimensions:  they calculate the distance between their
			data points and their relation, normalized by the errorbar.  While the errors on the \lya\ are modest, those on $X_\mathrm{LIS}^\mathrm{O3}$ are often large, spanning typical ranges of 0.3 dex on an axis that itself has a dynamic range of 2 dex. Were the errorbars larger still, their definition would report that the relation captures \emph{more} of the variance, and the reported quality of their relation would improve even though the ability of the data to discriminate between models has deteriorated.

			Secondly, they determine this value of 90\% by estimating the intrinsic
			scatter in their data ($\sigma_\mathrm{int}$) using the assumption that the total variance  can be written as the simple sum of variances: $\sigma_\mathrm{tot}^2 = \sigma_\mathrm{mod}^2 + \sigma_\mathrm{int}^2 + \sigma_\mathrm{meas}^2$, where $\sigma_\mathrm{tot}$ is the total dispersion, and $\sigma_\mathrm{mod}$, and $\sigma_\mathrm{meas}$ are the errors on the model and measurement, respectively.  This summation assumes that each term is both independent and Gaussianly distributed.  
			Additionally, this method, which re-samples data-points symmetrically around a model, requires that the real data are themselves distributed symmetrically.  Figure~7 of \citet{trainor2019} however shows an uneven distribution of points around the fitted model, which hints at an additional source of systematic uncertainty. 

			Nevertheless, the relation they present shows strong predictive potential and an exploration of how it performs at low-$z$ is a very interesting topic for future work.

    	\subsection{Potential uses of these relations}
			\subsubsection{Simulating the \lya\ luminosity function}
				
				Given the possible uses of \lya\ to probe evolving galaxies at high-$z$, numerous studies have attempted to model the \lya\ luminosity function (LF).  The approach almost always begins by generating a population of mock galaxies using either prescriptive (semi-)analytical models \citep{mao2007,dayal2008,ledelliou2005,ledelliou2006,orsi2008,kobayashi2010,garel2012,garel2015}, 
				or hydrodynamical models set in the cosmological framework \citep{nagamine2010,shimizu2011,dayal2012,hutter2015}.
				Then the \lya\ luminosity must be somehow assigned to each galaxy in the simulation, and their distribution converted into that which can be recovered by observation.  While many simulations can accurately model the UV LFs (say, of LBGs) this is more challenging for LAEs as none of the simulations contain sufficient information to estimate or accurately simulate the \lya. This is consequently done in a number of ways, such as assuming pure dust obscuration models \citep{kobayashi2010,shimizu2011}
				or assuming or fitting a constant escape fraction or duty cycle \citep[scaling \lstar\ and/or \phistar, respectively;][]{ledelliou2005,ledelliou2006,nagamine2010}
				Perhaps the most realistic effort has been presented by \citet{garel2012,garel2015} who used the theoretical grid of transfer models \citep{schaerer2011} to derive the \lya\ escape fraction galaxy-by-galaxy by looking up the object from the grid with the closest matching dust content, outflow velocity, gas column, etc.  However, even this approach is limited by the assumption of homogenous spherical shells of gas.

				The work presented in this paper offers a set of alternatives, that bypass many of the assumptions that enter the above models.  One may proceed in a manner similar to Garel et al, but instead of assigning \lya\ escape fraction from RT models, the \lya\ luminosity may be computed from all the other model parameters.  Indeed it is for this very reason that present our prediction results in physical properties (mass, SFR, etc) as well as observables.  While the limitation that the relation is derived from low-z galaxies will always remain, the resulting relation between \lya\ luminosity or escape fraction is likely to be a major improvement beyond assuming models of pure dust attenuation, constant escape fractions, or even modeling the RT using spherical shells.  It’s plausible, moreover, that these relations may be useful to predicting the \lya\ output of galaxies at the highest redshifts as discussed in the following section.

    		\subsubsection{Reionization studies}
    			One of the major possible applications of the relation we have derived is mapping the progress of reionization. As we look to higher redshifts a higher neutral fraction of the IGM translates to more \lya{} emitted from star forming galaxies being trapped and scattered out of our line of sight \citep{stark2010,stark2011,ono2012,mason2018,debarros2017,jensen2014}. If we do not know how much \lya{} we expect from a given galaxy at a given redshift it is impossible to determine how much of the \lya{} has been lost in the IGM. Our relation gives a possible way to predict the expected \lya{} output of a single galaxy. We can remove the uncertainty of the ISM transfer and galaxy evolution effects from the estimates of the transmitted fraction of \lya{} from many different galaxies and sightlines, potentially allowing us to probe the structure of the IGM towards the end of the EoR. 

    			Translating transmitted \lya{} fraction into exact neutral fraction encountered is not quite trivial however, since the opacity of neutral gas to \lya{} depends sensitively on the frequency shift of \lya{} from line center. For this reason we attempted to construct a predictive relation for the \lya{} red peak velocity (Section~\ref{sec: spectral properties}) but we found that our sample size is currently too small to extract a robust relation. The COS archive contains many observations where this can be measured and exploring this further is therefore a promising avenue for future work.

				A complication that arises when attempting to apply our results, or the results of \citet{trainor2019}, to high redshift galaxies is that some of the required observations will remain extremely challenging, even in the ELT era. In particular this applies to the properties of the LIS absorption lines. The Trainor et al study demanded $S/N>20$ in a stacked combination of 7 LIS lines which equates to approximately $S/N>8$ per line.  While conceivable for lensed galaxies at $z\approx 7$ it will be very demanding to obtain these data for large samples of normal galaxies in the field.  The relations presented here include the same absorption lines but they are also significantly lower in the ranking (Fig~\ref{fig: Forward selection distribution}) suggesting that similar predictive power can be obtained without such high spectroscopic SNR in the UV continuum.

    		\subsubsection{Intensity mapping}
    		Relatively recently there arisen a new and interesting alternative method for studying reionization known as intensity mapping. The idea is that instead of focusing on resolved individual sources, one attempts to measure the low surface brightness emission of all the galaxies in an area. By comparing the power spectrum of the spatial intensity fluctuations in the observations to those derived from simulations one can place constraints on reionization scenarios.

			However, in order to derive the expected \lya\ distribution from simulations one either has to do radiative transfer simulations over cosmologically relevant simulation sizes, but this requires a resolution which is not currently computationally achievable, or one has to apply some scaling relation that translates physical galaxy properties into an expected \lya\ luminosity.

			It is common to use scaling relations that assume that star-formation translates directly into \lya\ production with some assumed efficiency factor and that some fraction, $f_\mathrm{esc}$ of these photons escape the galaxy (see for instance equation 8 in \citet{silva2013} or equation 16 in \citet{comaschi2016}). Another approach is to calculate expected \lya\ structures directly from luminosity function extrapolations (e.g. \citealt{pullen2014}). This does carry with it strong assumptions, especially at high redshift (z$\geq$8) where the luminosity function of \lya\ emitters if poorly constrained.

			The work we have presented here provides another solution. Since we have been able to empirically map the physical properties of our galaxies to their total \lya\ luminosity we provide a relation that can be easily applied to link the physical properties derived from simulations to the expected \lya\ output. Using the relation we provide here should therefore provide a completely empirical mapping that can be easily applied to simulations with almost no computational cost.

\subsection{Outlook}
		Despite the long history of studies of \lya\ the multivariate approach to \lya\ prediction is still in its infancy. This is primarily due to the large time investments required to gather large datasets spanning many observables. Therefore there are still significant unexplored possibilities both in terms of the data included in the multivariate analysis and the specifics of the methodology used. This is well illustrated by the difference between the methods presented here and those of \citet{trainor2019} and \citet{yang2017} even though all these works result in strong predictive relations.

		The methodology we employ in this work is to fit a relatively simple unbiased powerlaw relation to a large variable set whereas \citet{trainor2019} and \citet{yang2017} construct their relations partially based on theoretical expectations of variable importance and physical reasoning. It is however possible to take a middle road between these two approaches. One could for instance use radiative transfer simulations to constrain expected the functional form of the \lya-to-variable-relations instead of simply assuming a power law or linear relation like we do in this work.

		From the point of view of the available data there are several ways pathways for future improvement. The simplest is naturally to increase the number of galaxies in the sample, preferentially by including ones that expand the covered parameter space i.e. galaxies that lie at the extremes of our current sample distribution. This would help create a more robust relation fit, more well determined variable importance rankings, and also increase the number of galaxies to which the relation can be meaningfully applied. Inclusion of variables more directly related to \lya\ escape, such as \hI\ column density measurements from higher order Lyman series lines, or resolved \hI\ 21 cm data such as that from \citet{pardy2014} and Le Reste et al. (in prep)  could also help provide a stronger relation.

		It may also be of value to look more closely into which variables are considered for the creation of a predictive relation. \citet{henry2018} showed a tight correlation between the escape fractions of \lya\ and the Mg~{\sc ii}~$\lambda\lambda 2796$,2803~\AA\ doublet, revealing a relationship that corresponds with 1:1 within the errorbars (their Fig.~10 and Eq.~5).  If observations can determine all the quantities required to measure $f_\mathrm{esc}$(MgII), and the SFR can independently measured, then the \lya\ luminosity could follow directly; this appears to be a more straightforward method compared to the methods we propose here.  However these methods also require measurements of the [O~{\sc ii}] and [O~{\sc iii}] to determine the intrinsic Mg~{\sc ii}, as well as (for example) \halpha\ and \hbeta\ to derive the SFR, and also a suite of photoionization models. Hence despite the physical simplicity of the model, it still requires a substantial set of observations in order to be realized. Further observations in a larger sample will determine how well this relationship will hold in a more diverse galaxy sample.

		There are several other parameter spaces that have not been thoroughly explored in this work. We have only used simplistic morphological descriptions in the current work but a more sophisticated treatment of the morphology of the galaxies,  such as the Gini coefficient, M20 or Petrosian radius, could provide more information. This has been explored for the 14 original LARS galaxies by \citet{guaita2015} and there they showed that the galaxies with strong \lya\ emission also showed more compact morphologies. This is also reflected in the high importance found for the UV compactness in this work. IFU observations also open up the possibility of studying the kinematics of the gas in the galaxy in detail. In this study we used the velocity shift and width of the low ionization lines as the only kinematical diagnostics but with resolved IFU data one can determine more detailed parameters such as the shear velocity and the resolved velocity dispersion. \citet{herenz2016} showed that for the LARS sample \lya\ emission was clearly correlated to how dispersion dominated the system is, i.e. how unordered the galaxy is. \citet{martin2015} also show the importance of kinematics for \lya\ escape, highlighting the role of fast outflows for creating escape channels for \lya\ in dusty galaxies.

\section{Conclusions}
	In this paper we made use of the LARS and extended LARS (eLARS) datasets which comprise detailed observations of 42 low redshift galaxies to show that \lya\ luminosity can be predicted from other galaxy properties, both direct observables (RMS=0.19, $R^2=0.93$) and derived physical properties (RMS=0.27 dex, $R^2=0.85$). We used statistical cross validation to determine that both of these relations appear to be very robust and generalize well to prediction of other \lya-emitting galaxy samples. For the relation based on observables we  were able to confirm this by accurately predicting the \lya\ luminosity of a separate set of compact starburst galaxies. 

	We also showed that it is possible to determine which variables made the most important contributions to the predictions. We find that the ranking of physical variables is highly robust and also agrees with theoretical expectations with SFR which is strongly related to \lya\ production, being the  most important and E$_{B-V}$, related to suppression of \lya\ escape, being the second most important. When using direct observables we showed that the strongest predictor of \lya\ luminosity is the Far UV luminosity.

	We therefore decided to fold out the first order predictors, SFR and FUV, and see if it was possible to predict the escape fraction and equivalent width of \lya\ respectively. Our results clearly demonstrate that it is possible to do so, but that the scatter on such relations is higher, as expected.

	The predictive relations derived here have important applications for for instance intensity mapping and probing the neutral fraction of the ISM. The variable importances we have derived also provide a clear guide for how future observations with JWST and the ELT should be prioritized.	
	The most important variables are the H$\alpha$ luminosity (SFR) for the physical variables and the FUV luminosity for the observable variables. The restframe FUV luminosity of galaxies is routinely detected with the HST at high redshifts, but \halpha\ will require JWST. Simple calculations using the JWST ETC show that JWST NIRSPEC will be able to detect the H$\alpha$ emission from a galaxy with SFR = 10 M$_\odot$yr$^{-1}$ at redshift 6 with a signal to noise of 10 per pixel in a little more than one hour of integration.

	Detecting the UV ISM lines with sufficient signal-to-noise to perform the kinematic measurements that we use in this study is more challenging however and will not be feasible even with the JWST. We therefore consider simulations of the MOSAIC multi-object-spectrograph on the ELT by \citet{disseau2014} and also presented in \citet{evans2015}. They show that UV ISM absorption lines  of a redshift 7, $J_{AB}=$26 mag galaxy, which corresponds to a star formation rate of roughly 10 M$_\odot$yr$^{-1}$, will be detectable with a signal-to-noise of 10 per pixel with 40 hours of integration time. The expensiveness of this observation will be somewhat mitigated by using the multiplexing capabilities of MOSAIC but it still requires a substantial time investment. Fortunately our variable selection has shown that these properties are not crucial for our prediction and we can construct a relation that is significantly cheaper to observe without losing much of the predictive power.

\section{Acknowledgements}
	M.H. acknowledges the support of the Swedish Research Council (Vetenskapsrådet)and the Swedish National Space Board (SNSB), and is Fellow of the Knut and Alice Wallenberg Foundation. D.K. is supported by the Centre National d’Etudes Spatiales (CNES)/Centre National de la Recherche Scientifique (CNRS), convention 131425

	We thank the referee for helpful comments that significantly improved the manuscript. We are grateful  for the useful feedback provided by Jorryt Matthee, Ryan Trainor, Thibault Garel, José Fonseca and E.C. Herenz. We also thank the Statistical Research Group at Stockholm University for their assistance with the statistical background for this paper.


\bibliography{Paper_I}

\begin{thebibliography}{}
\expandafter\ifx\csname natexlab\endcsname\relax\def\natexlab#1{#1}\fi
\providecommand{\url}[1]{\href{#1}{#1}}

\bibitem[{Adams(1972)}]{adams1972}
Adams, T.~F. 1972, The Astrophysical Journal, 174, 439

\bibitem[{Avila {et~al.}(2019)Avila, Bohlin, Hathi, Lockwood, \&
  Lim}]{avila2019}
Avila, R.~J., Bohlin, R., Hathi, N., Lockwood, S., \& Lim, P.~L. 2019, {{SBC
  Absolute Flux Calibration}}, Technical report

\bibitem[{Cardamone {et~al.}(2009)Cardamone, Schawinski, Sarzi, Bamford,
  Bennert, Urry, Lintott, Keel, Parejko, Nichol, Thomas, Andreescu, Murray,
  Raddick, Slosar, Szalay, \& VandenBerg}]{cardamone2009}
Cardamone, C., Schawinski, K., Sarzi, M., {et~al.} 2009, Monthly Notices of the
  Royal Astronomical Society, 399, 1191

\bibitem[{Cardelli {et~al.}(1989)Cardelli, Clayton, \& Mathis}]{cardelli1989}
Cardelli, J.~A., Clayton, G.~C., \& Mathis, J.~S. 1989, The Astrophysical
  Journal, 345, 245

\bibitem[{Caruana {et~al.}(2012)Caruana, Bunker, Wilkins, Stanway, Lacy,
  Jarvis, Lorenzoni, \& Hickey}]{caruana2012}
Caruana, J., Bunker, A.~J., Wilkins, S.~M., {et~al.} 2012, Monthly Notices of
  the Royal Astronomical Society, 427, 3055

\bibitem[{Caruana {et~al.}(2014)Caruana, Bunker, Wilkins, Stanway, Lorenzoni,
  Jarvis, \& Ebert}]{caruana2014}
---. 2014, Monthly Notices of the Royal Astronomical Society, 443, 2831

\bibitem[{Comaschi \& Ferrara(2016)}]{comaschi2016}
Comaschi, P., \& Ferrara, A. 2016, Monthly Notices of the Royal Astronomical
  Society, 455, 725

\bibitem[{Dayal \& Ferrara(2012)}]{dayal2012}
Dayal, P., \& Ferrara, A. 2012, Monthly Notices of the Royal Astronomical
  Society, 421, 2568

\bibitem[{Dayal {et~al.}(2008)Dayal, Ferrara, \& Gallerani}]{dayal2008}
Dayal, P., Ferrara, A., \& Gallerani, S. 2008, Monthly Notices of the Royal
  Astronomical Society, 389, 1683

\bibitem[{De~Barros {et~al.}(2017)De~Barros, Pentericci, Vanzella, Castellano,
  Fontana, Grazian, Conselice, Yan, Koekemoer, Cristiani, Dickinson,
  Finkelstein, \& Maiolino}]{debarros2017}
De~Barros, S., Pentericci, L., Vanzella, E., {et~al.} 2017, Astronomy \&
  Astrophysics, 608, A123

\bibitem[{Dijkstra(2014)}]{dijkstra2014}
Dijkstra, M. 2014, Publications of the Astronomical Society of Australia, 31,
  e040

\bibitem[{Dijkstra {et~al.}(2007)Dijkstra, Wyithe, \& Haiman}]{dijkstra2007}
Dijkstra, M., Wyithe, J. S.~B., \& Haiman, Z. 2007, Monthly Notices of the
  Royal Astronomical Society, 379, 253

\bibitem[{Disseau {et~al.}(2014)Disseau, Puech, Flores, Hammer, Yang, \&
  Pentericci}]{disseau2014}
Disseau, K., Puech, M., Flores, H., {et~al.} 2014, in {{SPIE Astronomical
  Telescopes}} + {{Instrumentation}}, ed. S.~K. Ramsay, I.~S. McLean, \&
  H.~Takami, {Montr{\'e}al, Quebec, Canada}, 914791

\bibitem[{Duval {et~al.}(2014)Duval, Schaerer, {\"O}stlin, \&
  Laursen}]{duval2014}
Duval, F., Schaerer, D., {\"O}stlin, G., \& Laursen, P. 2014, Astronomy \&
  Astrophysics, 562, A52

\bibitem[{Evans {et~al.}(2015)Evans, Puech, Afonso, Almaini, Amram, Aussel,
  Barbuy, Basden, Bastian, Battaglia, Biller, Bonifacio, Bouch{\'e}, Bunker,
  Caffau, Charlot, Cirasuolo, Clenet, Combes, Conselice, Contini, Cuby, Dalton,
  Davies, {de Koter}, Disseau, Dunlop, Epinat, Fiore, Feltzing, Ferguson,
  Flores, Fontana, Fusco, Gadotti, Gallazzi, Gallego, Giallongo, Gon{\c
  c}alves, Gratadour, Guenther, Hammer, Hill, {Huertas-Company}, Ibata, Kaper,
  Korn, Larsen, F{\`e}vre, Lemasle, Maraston, Mei, Mellier, Morris, {\"O}stlin,
  Paumard, Pello, Pentericci, Peroux, Petitjean, Rodrigues,
  {Rodr{\'i}guez-Mu{\~n}oz}, Rouan, Sana, Schaerer, Telles, Trager, Tresse,
  Welikala, Zibetti, \& Ziegler}]{evans2015}
Evans, C., Puech, M., Afonso, J., {et~al.} 2015, arXiv:1501.04726 [astro-ph],
  arXiv:1501.04726

\bibitem[{Fontana {et~al.}(2010)Fontana, Vanzella, Pentericci, Castellano,
  Giavalisco, Grazian, Boutsia, Cristiani, Dickinson, Giallongo, Maiolino,
  Moorwood, \& Santini}]{fontana2010}
Fontana, A., Vanzella, E., Pentericci, L., {et~al.} 2010, The Astrophysical
  Journal, 725, L205

\bibitem[{Garel {et~al.}(2015)Garel, Blaizot, Guiderdoni, {Michel-Dansac},
  Hayes, \& Verhamme}]{garel2015}
Garel, T., Blaizot, J., Guiderdoni, B., {et~al.} 2015, Monthly Notices of the
  Royal Astronomical Society, 450, 1279

\bibitem[{Garel {et~al.}(2012)Garel, Blaizot, Guiderdoni, Schaerer, Verhamme,
  \& Hayes}]{garel2012}
---. 2012, Monthly Notices of the Royal Astronomical Society, 422, 310

\bibitem[{Guaita {et~al.}(2015)Guaita, Melinder, Hayes, {\"O}stlin, Gonzalez,
  Micheva, Adamo, {Mas-Hesse}, Sandberg, {Ot{\'i}-Floranes}, Schaerer,
  Verhamme, Freeland, Orlitov{\'a}, Laursen, Cannon, Duval, {Rivera-Thorsen},
  Herenz, Kunth, Atek, Puschnig, Gruyters, \& Pardy}]{guaita2015}
Guaita, L., Melinder, J., Hayes, M., {et~al.} 2015, Astronomy \& Astrophysics,
  576, A51

\bibitem[{Haiman \& Spaans(1999)}]{haiman1999}
Haiman, Z., \& Spaans, M. 1999, The Astrophysical Journal, 518, 138

\bibitem[{Hayes(2015)}]{hayes2015}
Hayes, M. 2015, Publications of the Astronomical Society of Australia, 32, e027

\bibitem[{Hayes {et~al.}(2009)Hayes, {\"O}stlin, {Mas-Hesse}, \&
  Kunth}]{hayes2009}
Hayes, M., {\"O}stlin, G., {Mas-Hesse}, J.~M., \& Kunth, D. 2009, The
  Astronomical Journal, 138, 911

\bibitem[{Hayes {et~al.}(2011)Hayes, Schaerer, {\"O}stlin, {Mas-Hesse}, Atek,
  \& Kunth}]{hayes2011}
Hayes, M., Schaerer, D., {\"O}stlin, G., {et~al.} 2011, The Astrophysical
  Journal, 730, 8

\bibitem[{Hayes {et~al.}(2013)Hayes, {\"O}stlin, Schaerer, Verhamme,
  {Mas-Hesse}, Adamo, Atek, Cannon, Duval, Guaita, Herenz, Kunth, Laursen,
  Melinder, Orlitov{\'a}, {Ot{\'i}-Floranes}, \& Sandberg}]{hayes2013}
Hayes, M., {\"O}stlin, G., Schaerer, D., {et~al.} 2013, The Astrophysical
  Journal, 765, L27

\bibitem[{Hayes {et~al.}(2014)Hayes, {\"O}stlin, Duval, Sandberg, Guaita,
  Melinder, Adamo, Schaerer, Verhamme, Orlitov{\'a}, {Mas-Hesse}, Cannon, Atek,
  Kunth, Laursen, {Ot{\'i}-Floranes}, Pardy, {Rivera-Thorsen}, \&
  Herenz}]{hayes2014}
Hayes, M., {\"O}stlin, G., Duval, F., {et~al.} 2014, The Astrophysical Journal,
  782, 6

\bibitem[{Heckman {et~al.}(2011)Heckman, Borthakur, Overzier, Kauffmann,
  {Basu-Zych}, Leitherer, Sembach, Martin, Rich, Schiminovich, \&
  Seibert}]{heckman2011}
Heckman, T.~M., Borthakur, S., Overzier, R., {et~al.} 2011, The Astrophysical
  Journal, 730, 5

\bibitem[{Henry {et~al.}(2018)Henry, Berg, Scarlata, Verhamme, \&
  Erb}]{henry2018}
Henry, A., Berg, D.~A., Scarlata, C., Verhamme, A., \& Erb, D. 2018, The
  Astrophysical Journal, 855, 96

\bibitem[{Henry {et~al.}(2015)Henry, Scarlata, Martin, \& Erb}]{henry2015}
Henry, A., Scarlata, C., Martin, C.~L., \& Erb, D. 2015, The Astrophysical
  Journal, 809, 19

\bibitem[{Herenz {et~al.}(2016)Herenz, Gruyters, Orlitova, Hayes, {\"O}stlin,
  Cannon, Roth, Bik, Pardy, {Ot{\'i}-Floranes}, {Miguel Mas-Hesse}, Adamo,
  Atek, Duval, Guaita, Kunth, Laursen, Melinder, Puschnig, {Rivera-Thorsen},
  Schaerer, \& Verhamme}]{herenz2016}
Herenz, E.~C., Gruyters, P., Orlitova, I., {et~al.} 2016, Astronomy \&
  Astrophysics, 587, A78

\bibitem[{Hutter {et~al.}(2015)Hutter, Dayal, \& M{\"u}ller}]{hutter2015}
Hutter, A., Dayal, P., \& M{\"u}ller, V. 2015, Monthly Notices of the Royal
  Astronomical Society, 450, 4025

\bibitem[{Izotov {et~al.}(2011)Izotov, Guseva, \& Thuan}]{izotov2011}
Izotov, Y.~I., Guseva, N.~G., \& Thuan, T.~X. 2011, The Astrophysical Journal,
  728, 161

\bibitem[{Izotov {et~al.}(2016)Izotov, Schaerer, Thuan, Worseck, Guseva,
  Orlitov{\'a}, \& Verhamme}]{izotov2016}
Izotov, Y.~I., Schaerer, D., Thuan, T.~X., {et~al.} 2016, Monthly Notices of
  the Royal Astronomical Society, 461, 3683

\bibitem[{Izotov {et~al.}(2018)Izotov, Schaerer, Worseck, Guseva, Thuan,
  Verhamme, Orlitov{\'a}, \& Fricke}]{izotov2018}
Izotov, Y.~I., Schaerer, D., Worseck, G., {et~al.} 2018, Monthly Notices of the
  Royal Astronomical Society, 474, 4514

\bibitem[{Jaskot \& Oey(2014)}]{jaskot2014}
Jaskot, A.~E., \& Oey, M.~S. 2014, The Astrophysical Journal, 791, L19

\bibitem[{Jensen {et~al.}(2014)Jensen, Hayes, Iliev, Laursen, Mellema, \&
  Zackrisson}]{jensen2014}
Jensen, H., Hayes, M., Iliev, I.~T., {et~al.} 2014, Monthly Notices of the
  Royal Astronomical Society, 444, 2114

\bibitem[{Jung {et~al.}(2018)Jung, Finkelstein, Livermore, Dickinson, Larson,
  Papovich, Song, Tilvi, \& Wold}]{jung2018}
Jung, I., Finkelstein, S.~L., Livermore, R.~C., {et~al.} 2018, The
  Astrophysical Journal, 864, 103

\bibitem[{Kashikawa {et~al.}(2006)Kashikawa, Shimasaku, Malkan, Doi, Matsuda,
  Ouchi, Taniguchi, Ly, Nagao, Iye, Motohara, Murayama, Murozono, Nariai, Ohta,
  Okamura, Sasaki, Shioya, \& Umemura}]{kashikawa2006}
Kashikawa, N., Shimasaku, K., Malkan, M.~A., {et~al.} 2006, The Astrophysical
  Journal, 648, 7

\bibitem[{Kennicutt(1998)}]{kennicuttjr.1998}
Kennicutt, Jr., R.~C. 1998, The Astrophysical Journal, 498, 541

\bibitem[{Kewley \& Dopita(2002)}]{kewley2002}
Kewley, L.~J., \& Dopita, M.~A. 2002, The Astrophysical Journal Supplement
  Series, 142, 35

\bibitem[{Kobayashi {et~al.}(2010)Kobayashi, Totani, \&
  Nagashima}]{kobayashi2010}
Kobayashi, M. A.~R., Totani, T., \& Nagashima, M. 2010, The Astrophysical
  Journal, 708, 1119

\bibitem[{Kobulnicky \& Kewley(2004)}]{kobulnicky2004}
Kobulnicky, H.~A., \& Kewley, L.~J. 2004, The Astrophysical Journal, 617, 240

\bibitem[{Kunth {et~al.}(1998)Kunth, {Mas-Hesse}, Terlevich, Terlevich,
  Lequeux, \& Fall}]{kunth1998}
Kunth, D., {Mas-Hesse}, J.~M., Terlevich, E., {et~al.} 1998, Astronomy and
  Astrophysics, 334, 11

\bibitem[{Laursen {et~al.}(2013)Laursen, Duval, \& {\"O}stlin}]{laursen2013a}
Laursen, P., Duval, F., \& {\"O}stlin, G. 2013, The Astrophysical Journal, 766,
  124

\bibitem[{Le~Delliou {et~al.}(2005)Le~Delliou, Lacey, Baugh, Guiderdoni, Bacon,
  Courtois, Sousbie, \& Morris}]{ledelliou2005}
Le~Delliou, M., Lacey, C., Baugh, C.~M., {et~al.} 2005, Monthly Notices of the
  Royal Astronomical Society: Letters, 357, L11

\bibitem[{Le~Delliou {et~al.}(2006)Le~Delliou, Lacey, Baugh, \&
  Morris}]{ledelliou2006}
Le~Delliou, M., Lacey, C.~G., Baugh, C.~M., \& Morris, S.~L. 2006, Monthly
  Notices of the Royal Astronomical Society, 365, 712

\bibitem[{Malhotra \& Rhoads(2004)}]{malhotra2004}
Malhotra, S., \& Rhoads, J.~E. 2004, The Astrophysical Journal, 617, L5

\bibitem[{Malhotra \& Rhoads(2006)}]{malhotra2006}
---. 2006, The Astrophysical Journal, 647, L95

\bibitem[{Mao {et~al.}(2007)Mao, Lapi, Granato, {de Zotti}, \&
  Danese}]{mao2007}
Mao, J., Lapi, A., Granato, G.~L., {de Zotti}, G., \& Danese, L. 2007, The
  Astrophysical Journal, 667, 655

\bibitem[{Martin {et~al.}(2015)Martin, Dijkstra, Henry, Soto, Danforth, \&
  Wong}]{martin2015}
Martin, C.~L., Dijkstra, M., Henry, A., {et~al.} 2015, The Astrophysical
  Journal, 803, 6

\bibitem[{Mason {et~al.}(2018{\natexlab{a}})Mason, Treu, Dijkstra, Mesinger,
  Trenti, Pentericci, {de Barros}, \& Vanzella}]{mason2018a}
Mason, C.~A., Treu, T., Dijkstra, M., {et~al.} 2018{\natexlab{a}}, The
  Astrophysical Journal, 856, 2

\bibitem[{Mason {et~al.}(2018{\natexlab{b}})Mason, Treu, {de Barros}, Dijkstra,
  Fontana, Mesinger, Pentericci, Trenti, \& Vanzella}]{mason2018}
Mason, C.~A., Treu, T., {de Barros}, S., {et~al.} 2018{\natexlab{b}}, The
  Astrophysical Journal, 857, L11

\bibitem[{Mason {et~al.}(2019)Mason, Fontana, Treu, Schmidt, Hoag, Abramson,
  Amorin, Brada{\v c}, Guaita, Jones, Henry, Malkan, Pentericci, Trenti, \&
  Vanzella}]{mason2019}
Mason, C.~A., Fontana, A., Treu, T., {et~al.} 2019, Monthly Notices of the
  Royal Astronomical Society, 485, 3947

\bibitem[{{Morice-Atkinson} {et~al.}(2018){Morice-Atkinson}, Hoyle, \&
  Bacon}]{morice-atkinson2018}
{Morice-Atkinson}, X., Hoyle, B., \& Bacon, D. 2018, Monthly Notices of the
  Royal Astronomical Society, 481, 4194

\bibitem[{Nagamine {et~al.}(2010)Nagamine, Ouchi, Springel, \&
  Hernquist}]{nagamine2010}
Nagamine, K., Ouchi, M., Springel, V., \& Hernquist, L. 2010, Publications of
  the Astronomical Society of Japan, 62, 1455

\bibitem[{Neufeld(1990)}]{neufeld1990}
Neufeld, D.~A. 1990, The Astrophysical Journal, 350, 216

\bibitem[{Ono {et~al.}(2012)Ono, Ouchi, Mobasher, Dickinson, Penner, Shimasaku,
  Weiner, Kartaltepe, Nakajima, Nayyeri, Stern, Kashikawa, \&
  Spinrad}]{ono2012}
Ono, Y., Ouchi, M., Mobasher, B., {et~al.} 2012, The Astrophysical Journal,
  744, 83

\bibitem[{Orsi {et~al.}(2008)Orsi, Lacey, Baugh, \& Infante}]{orsi2008}
Orsi, A., Lacey, C.~G., Baugh, C.~M., \& Infante, L. 2008, Monthly Notices of
  the Royal Astronomical Society, 391, 1589

\bibitem[{Osterbrock(1962)}]{osterbrock1962}
Osterbrock, D.~E. 1962, The Astrophysical Journal, 135, 195

\bibitem[{{\"O}stlin {et~al.}(2014){\"O}stlin, Hayes, Duval, Sandberg,
  {Rivera-Thorsen}, Marquart, Orlitov{\'a}, Adamo, Melinder, Guaita, Atek,
  Cannon, Gruyters, Herenz, Kunth, Laursen, {Mas-Hesse}, Micheva,
  {Ot{\'i}-Floranes}, Pardy, Roth, Schaerer, \& Verhamme}]{ostlin2014}
{\"O}stlin, G., Hayes, M., Duval, F., {et~al.} 2014, The Astrophysical Journal,
  797, 11

\bibitem[{Pardy {et~al.}(2014)Pardy, Cannon, {\"O}stlin, Hayes,
  {Rivera-Thorsen}, Sandberg, Adamo, Freeland, Herenz, Guaita, Kunth, Laursen,
  {Mas-Hesse}, Melinder, Orlitov{\'a}, {Ot{\'i}-Floranes}, Puschnig, Schaerer,
  \& Verhamme}]{pardy2014}
Pardy, S.~A., Cannon, J.~M., {\"O}stlin, G., {et~al.} 2014, The Astrophysical
  Journal, 794, 101

\bibitem[{Pedregosa {et~al.}(2011)Pedregosa, Varoquaux, Gramfort, Michel,
  Thirion, Grisel, Blondel, Prettenhofer, Weiss, Dubourg, Vanderplas, Passos,
  \& Cournapeau}]{pedregosa2011}
Pedregosa, F., Varoquaux, G., Gramfort, A., {et~al.} 2011, Journal of Machine
  Learning Research, 12, 2825

\bibitem[{Peng {et~al.}(2002)Peng, Ho, Impey, \& Rix}]{peng2002}
Peng, C.~Y., Ho, L.~C., Impey, C.~D., \& Rix, H.-W. 2002, The Astronomical
  Journal, 124, 266

\bibitem[{Pentericci {et~al.}(2011)Pentericci, Fontana, Vanzella, Castellano,
  Grazian, Dijkstra, Boutsia, Cristiani, Dickinson, Giallongo, Giavalisco,
  Maiolino, Moorwood, Paris, \& Santini}]{pentericci2011}
Pentericci, L., Fontana, A., Vanzella, E., {et~al.} 2011, The Astrophysical
  Journal, 743, 132

\bibitem[{Pullen {et~al.}(2014)Pullen, Dor{\'e}, \& Bock}]{pullen2014}
Pullen, A.~R., Dor{\'e}, O., \& Bock, J. 2014, The Astrophysical Journal, 786,
  111

\bibitem[{{Rivera-Thorsen} {et~al.}(2015){Rivera-Thorsen}, Hayes, {\"O}stlin,
  Duval, Orlitov{\'a}, Verhamme, {Mas-Hesse}, Schaerer, Cannon,
  {Ot{\'i}-Floranes}, Sandberg, Guaita, Adamo, Atek, Herenz, Kunth, Laursen, \&
  Melinder}]{rivera-thorsen2015}
{Rivera-Thorsen}, T.~E., Hayes, M., {\"O}stlin, G., {et~al.} 2015, The
  Astrophysical Journal, 805, 14

\bibitem[{Scarlata {et~al.}(2009)Scarlata, Colbert, Teplitz, Panagia, Hayes,
  Siana, Rau, Francis, Caon, Pizzella, \& Bridge}]{scarlata2009}
Scarlata, C., Colbert, J., Teplitz, H.~I., {et~al.} 2009, The Astrophysical
  Journal, 704, L98

\bibitem[{Schaerer {et~al.}(2011)Schaerer, Hayes, Verhamme, \&
  Teyssier}]{schaerer2011}
Schaerer, D., Hayes, M., Verhamme, A., \& Teyssier, R. 2011, Astronomy \&
  Astrophysics, 531, A12

\bibitem[{Schenker {et~al.}(2014)Schenker, Ellis, Konidaris, \&
  Stark}]{schenker2014}
Schenker, M.~A., Ellis, R.~S., Konidaris, N.~P., \& Stark, D.~P. 2014, The
  Astrophysical Journal, 795, 20

\bibitem[{Shimizu {et~al.}(2011)Shimizu, Yoshida, \& Okamoto}]{shimizu2011}
Shimizu, I., Yoshida, N., \& Okamoto, T. 2011, Monthly Notices of the Royal
  Astronomical Society, 418, 2273

\bibitem[{Silva {et~al.}(2013)Silva, Santos, Gong, Cooray, \& Bock}]{silva2013}
Silva, M.~B., Santos, M.~G., Gong, Y., Cooray, A., \& Bock, J. 2013, The
  Astrophysical Journal, 763, 132

\bibitem[{Stark {et~al.}(2010)Stark, Ellis, Chiu, Ouchi, \& Bunker}]{stark2010}
Stark, D.~P., Ellis, R.~S., Chiu, K., Ouchi, M., \& Bunker, A. 2010, Monthly
  Notices of the Royal Astronomical Society, 408, 1628

\bibitem[{Stark {et~al.}(2011)Stark, Ellis, \& Ouchi}]{stark2011}
Stark, D.~P., Ellis, R.~S., \& Ouchi, M. 2011, The Astrophysical Journal, 728,
  L2

\bibitem[{Trainor {et~al.}(2019)Trainor, Strom, Steidel, Rudie, Chen, \&
  Theios}]{trainor2019}
Trainor, R.~F., Strom, A.~L., Steidel, C.~C., {et~al.} 2019, arXiv:1908.04794
  [astro-ph], arXiv:1908.04794

\bibitem[{Treu {et~al.}(2012)Treu, Trenti, Stiavelli, Auger, \&
  Bradley}]{treu2012}
Treu, T., Trenti, M., Stiavelli, M., Auger, M.~W., \& Bradley, L.~D. 2012, The
  Astrophysical Journal, 747, 27

\bibitem[{Verhamme {et~al.}(2006)Verhamme, Schaerer, \&
  Maselli}]{verhamme2006a}
Verhamme, A., Schaerer, D., \& Maselli, A. 2006, Astronomy \& Astrophysics,
  460, 397

\bibitem[{Yang {et~al.}(2017)Yang, Malhotra, Gronke, Rhoads, Leitherer,
  Wofford, Jiang, Dijkstra, Tilvi, \& Wang}]{yang2017}
Yang, H., Malhotra, S., Gronke, M., {et~al.} 2017, The Astrophysical Journal,
  844, 171

\bibitem[{Yin {et~al.}(2007)Yin, Liang, Hammer, Brinchmann, Zhang, Deng, \&
  Flores}]{yin2007}
Yin, S.~Y., Liang, Y.~C., Hammer, F., {et~al.} 2007, Astronomy \& Astrophysics,
  462, 535

\end{thebibliography}

\end{document}